\documentclass[aps,prl,twocolumn,letterpaper,superscriptaddress,10pt]{revtex4-2}
\usepackage{amssymb,amsthm,amsmath,amsfonts}
\usepackage{graphicx,ulem,mathptmx}
\usepackage[shortlabels]{enumitem}
\usepackage[pdftex,dvipsnames,usenames]{xcolor}
\usepackage[colorlinks=true,urlcolor=blue,citecolor=blue,linkcolor=blue]{hyperref}

\begin{document}

\title{Tracking quantum coherence in polariton condensates with time-resolved tomography}

\author{Carolin L\"uders}
    \email{carolin.lueders@tu-dortmund.de}
	\affiliation{Experimentelle Physik 2, Technische Universit\"at Dortmund, D-44221 Dortmund, Germany}
\author{Matthias Pukrop}
	\affiliation{Department of Physics and Center for Optoelectronics and Photonics Paderborn (CeOPP), Universit\"at Paderborn, 33098 Paderborn, Germany}
\author{Franziska Barkhausen}
	\affiliation{Department of Physics and Center for Optoelectronics and Photonics Paderborn (CeOPP), Universit\"at Paderborn, 33098 Paderborn, Germany}
\author{Elena Rozas}
	\affiliation{Experimentelle Physik 2, Technische Universit\"at Dortmund, D-44221 Dortmund, Germany}
\author{Christian Schneider}
    \affiliation{Institute of Physics, University of Oldenburg, D-26129 Oldenburg, Germany}
\author{Sven H\"ofling}
    \affiliation{Technische Physik, Physikalisches Institut and W\"urzburg-Dresden Cluster of Excellence ct.qmat, Universit\"at W\"urzburg, 97074 W\"urzburg, Germany}
\author{Jan Sperling}
    \email{jan.sperling@upb.de}
    \affiliation{Theoretical Quantum Science, Institute for Photonic Quantum Systems (PhoQS), Paderborn University, Warburger Stra\ss{}e 100, 33098 Paderborn, Germany}
\author{Stefan Schumacher}
	\affiliation{Department of Physics and Center for Optoelectronics and Photonics Paderborn (CeOPP), Universit\"at Paderborn, 33098 Paderborn, Germany}
	\affiliation{Wyant College of Optical Sciences, University of Arizona, Tucson, Arizona 85721, USA}
\author{Marc A\ss{}mann}
	\affiliation{Experimentelle Physik 2, Technische Universit\"at Dortmund, D-44221 Dortmund, Germany}

\begin{abstract}
    Long-term quantum coherence constitutes one of the main challenges when engineering quantum devices.
    However, easily accessible means to quantify complex decoherence mechanisms are not readily available, nor are sufficiently stable systems.
    We harness novel phase-space methods---expressed through non-Gaussian convolutions of highly singular Glauber--Sudarshan quasiprobabilities---to dynamically monitor quantum coherence in polariton condensates with significantly enhanced coherence times.
    Via intensity- and time-resolved reconstructions of such phase-space functions from homodyne detection data, we probe the systems's resourcefulness for quantum information processing up to the nanosecond regime.
    Our experimental findings are confirmed through numerical simulations for which we develop an approach that renders established algorithms compatible with our methodology.
    In contrast to commonly applied phase-space functions, our distributions can be directly sampled from measured data, including uncertainties, and yield a simple operational measure of quantum coherence via the distribution's variance in phase.
    Therefore, we present a broadly applicable framework and a platform to explore time-dependent quantum phenomena and resources.
\end{abstract}

\date{\today}
\maketitle

\paragraph*{Introduction.---}\hspace*{-2.5ex}
	Applications in quantum information science require stable quantum superpositions as a key resource.
	Therefore, a great deal of effort is dedicated to quantifying quantum coherence \cite{SAP17,CG19,BCP14,LM14,SV15,WY16}.
	Beyond assessing the amount of useful quantumness, however, the evolution of quantum coherence is, at least, equally important to actually process information in quantum algorithms \cite{NC00,DM03,D20}.
	Furthermore, quantum coherence, providing a single number, does not yield an exhaustive quantum state description, generally not allowing for proposing schemes to access particular quantum resources.
	In this work, we overcome the challenging problem of studying dynamic quantum coherence, while also providing a comprehensive and accessible quantum state description.

    In quantum information science, quantum coherence is based on a set of computational and orthonormal basis states \cite{BCP14}, which are number states $|n\rangle$ in our study.
    Exceeding incoherent mixtures, i.e., diagonal $\hat\rho=\sum_{n}p_n|n\rangle\langle n|$, results in quantum superpositions, e.g., $\hat\rho=|\psi\rangle\langle\psi|$ with $|\psi\rangle=\sum_n \psi_n |n\rangle$, that form the foundation of quantum algorithms \cite{SV15}, such as quantum teleportation \cite{BBCJPW93}, Shor's factorization \cite{S94}, quantum key distribution \cite{BB84}, etc.
    It was shown in Ref. \cite{LPRSHSSA21} that, in optical and semiconductor systems, this operational quantification of quantum coherence for practical quantum protocols is distinctively different from commonly applied notions of nonclassicality and macroscopic coherence as characterized through negativities in phase-space functions \cite{SV20} and determined by correlation functions \cite{SV05}, respectively.

    Polaritons in semiconductor microcavities present a so-far untapped resource of quantum coherence \cite{LPRSHSSA21}.
    Polaritons---i.e., hybrid light-matter quasiparticles---arise from the strong coupling between the cavity photons and the quantum well excitons.
    Under nonresonant excitation, polaritons can spontaneously form a macroscopic coherent condensate \cite{DHY10,CC13}.
    The buildup of quantum coherence was demonstrated across the condensation threshold while an incoherent thermal behavior was observed below threshold \cite{LPRSHSSA21}.
    Nevertheless, the evolution and other vital information about the produced states were inaccessible, posing open problems to date.
    Generally, the dynamics of polariton condensates has been studied in previous experiments, mainly by measuring the field correlation $g^{(1)}$, using Michelson and Mach-Zehnder interferometry \cite{LKWBSRBSWAD08,KZWFBKSHD16,APAZLLL19,OTPSO21}, and the second-order correlation $g^{(2)}$ via Hanbury~Brown--Twiss interferometry \cite{KFASBAKSH18,BZSGTAL22}.
    But such correlation functions do not yield information about quantum coherence in the quantum-informational sense \cite{LPRSHSSA21}, which depends on superimposing computational basis states \cite{BCP14}.

    In this contribution, we establish a directly applicable and easily accessible method for analyzing quantum coherence via time-resolved quantum tomography.
    To this end, we adopt advanced phase-space functions, labeled as $P_\Omega$, for tracking quantum coherence produced by a state-of-the-art polariton microcavity system.
    The phase variance of $P_{\Omega}$ allows us to quantify quantum coherence, and we demonstrate that $P_{\Omega}$ can be directly sampled from our data using pattern functions.
    Exceptionally long coherence times---up to $1\,390~\mathrm{ps}$---are experimentally observed.
    This is further supported by microscopic simulations, advancing the commonly applied truncated Wigner approximation \cite{SLC02,BBDBG08} to phase-space methods based on $P_\Omega$.
    This renders it possible to reconstruct $P_{\Omega}$ directly, compare the numerical results to our experiment, reveal important parameter dynamics, and relate our measurements to key aspects of the underlying physical system.

\paragraph{Regularized phase-space functions $P_\Omega$---}\hspace*{-2.5ex}
    Phase-space distributions based on coherent states $|\alpha\rangle$ are useful for determining typical nonclassical effects, and they provide a full quantum state description \cite{SV20}.
    However, information-based quantum coherence relies on orthogonal as the classical reference, such as such as number states $|n\rangle$ but not coherent states.
    Moreover, the fundamental Glauber--Sudarshan phase-space distribution $P$ \cite{G63,S63}, where $\hat\rho=\int d^2\alpha\,P(\alpha)|\alpha\rangle\langle\alpha|$, can exhibit an exponential order of singularities \cite{S16}.
    And convolution-based regularizations were developed \cite{CG69,AW70},
    \begin{equation}
        P_\Omega(\alpha)=\int d^2\gamma\,\Omega(\gamma-\alpha)P(\gamma).
    \end{equation}
    For instance, the seminal Husimi $Q$ function and the Wigner function are obtained when $\Omega$ is a Gaussian kernel.
    However, the reconstruction of such common phase-space functions can be rather challenging, including ill-posed inversions \cite{T97,SSG09}, diverging pattern functions \cite{R96,LMKRR96}, and demanding maximum-likelihood estimations \cite{H97,L04,KWR04}.

    For optical scenarios, a regularization of $P$ distributions has been proposed to characterize nonclassical light by utilizing non-Gaussian kernels \cite{KV10};
    see Refs. \cite{KVTMS21,KASSVSH21} for recent experiments.
    While phase-space representations have been applied to semiconductor systems in pioneering papers \cite{KK08,KKS11,HLC14}, to date, this modern non-Gaussian approach to phase-space representations has not been exploited to analyze quantum coherence properties of semiconductors and its dynamics.

    Here, we use the non-Gaussian and phase-invariant kernel
	\begin{equation}
		\label{eq:ChosenKernel}
		\Omega(\gamma)=\left[\frac{J_1(2R|\gamma|)}{\sqrt{\pi}|\gamma|}\right]^2,
	\end{equation}
	with $J_1$ denoting the first Bessel function of the first kind and $R>0$ being an adjustable width parameter.
	Remarkably, the regularized function $P_\Omega$ can be directly obtained from data \cite{KVHS11}, with $P_\Omega(\alpha)\approx \sum_i w_i f_\Omega(\alpha,x_i;\varphi_i)$, circumventing all aforementioned reconstruction problems.
	In the formula, $(x_i,\varphi_i)$ is the $i$th quadrature measurement from balanced homodyne detection, $w_i$ are weights, and $f_\Omega$ are pattern functions;
	see Supplemental Material \cite{SM} for details.

    Incoherent contributions to the density operator, e.g., $|n\rangle\langle n|$, relate to phase-independent parts of $P_\Omega$, depending on the amplitude $|\alpha|$ only.
    Conversely, the distribution of the phase $\phi=\arg\alpha$ is directly connected to off-diagonal density operator terms, $|n\rangle\langle m|$ for $m\neq n$, thus determining quantum coherence \cite{LPRSHSSA21,SM}.
    Therefore, we here introduce the width of $P_\Omega$ in phase as an operational measure of quantum coherence.
    Specifically, the circular variance \cite{F93}
    \begin{equation}
        \label{eq:cric_var}
        \mathrm{Var}(\phi)=1-|\langle r\rangle|,
        \text{ with }
        \langle r\rangle=\int d^2\alpha\, P_\Omega(\alpha)\frac{\alpha}{|\alpha|},
    \end{equation}
    takes the maximal value one for a fully phase-randomized---i.e., incoherent---state;
    a narrow phase distribution---hence, a low circular variance---is obtained for strongly off-diagonal density operators, verifying a high degree of quantum coherence \cite{SM}.
    Therefore, the evolution of this circular variance of $P_\Omega$ monitors the dynamics of quantum coherence.

\begin{figure}
    \includegraphics[width=0.43\textwidth]{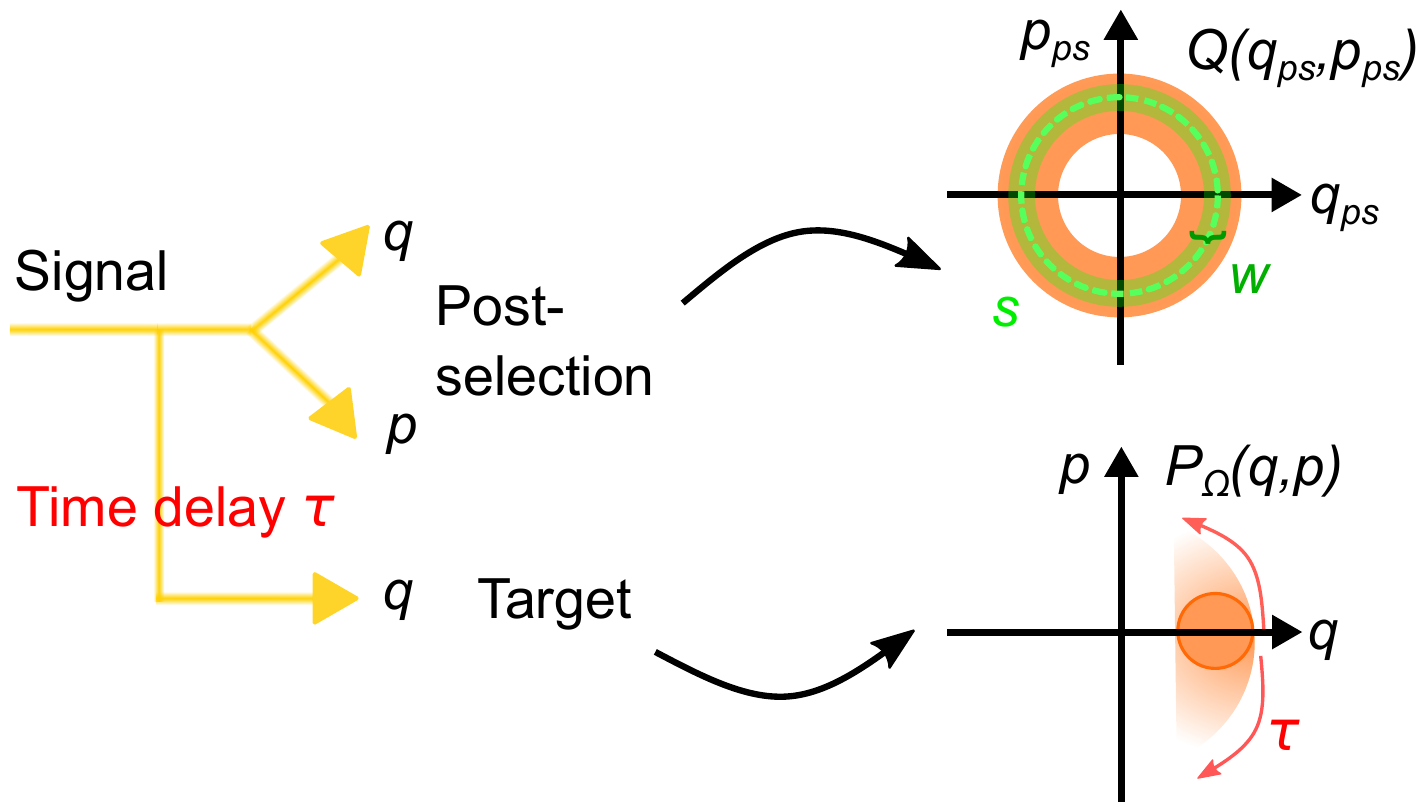}
    \caption{%
        Concept of the experiment as a conditional intensity spectroscopy.
        The signal is split into three homodyne detection channels.
        Two are used for intensity selection, delivering the field quadratures $(q_{ps}, p_{ps})$ of a Husimi function $Q(\alpha_{ps})$, with $\alpha_{ps}=q_{ps}+ip_{ps}$.
        Thereby, a specific region with radius $s$ and width $w$ is chosen, fixing the signal's mean intensity and its uncertainty.
        The third quadrature is the target channel for quadratures $q$.
        Together with the corresponding phases, this allows us to  directly sample $P_\Omega$.
        The time delay $\tau$ determines the elapsed time of the system's evolution.
    }\label{Fig:idea}
\end{figure}

\begin{figure*}
    \includegraphics[width=1\textwidth]{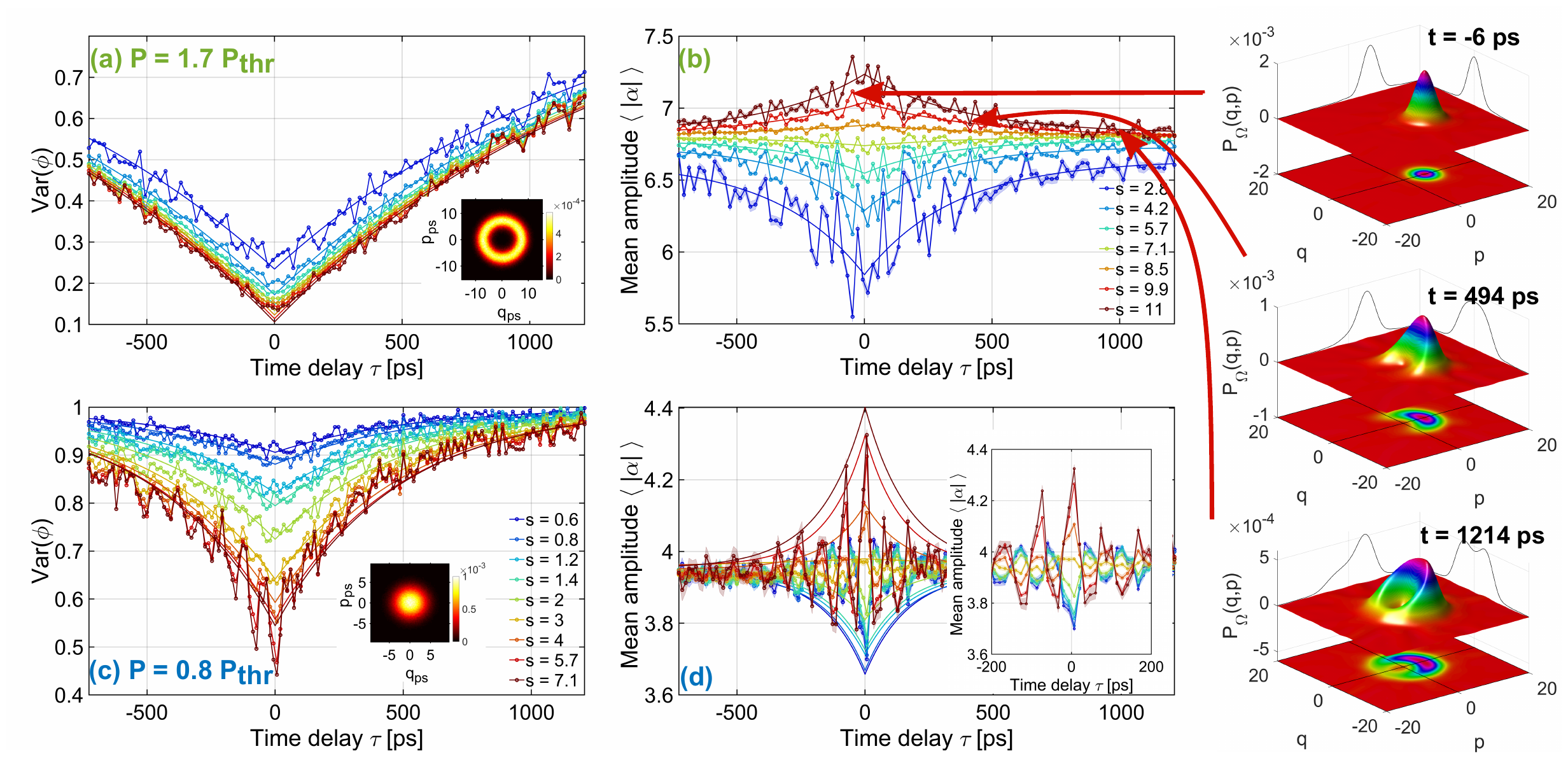}
    \caption{%
        Temporal behavior of the mean amplitude $\langle |\alpha|\rangle$ and the circular phase variance $\mathrm{Var}(\phi)$ of reconstructed $P_\Omega$ functions, depending on the selected intensity radius $s$.
        A shaded area around the curves corresponds to a one standard-deviation error margin, directly derived from $P_\Omega$'s uncertainty, but is mostly not visible.  
        The lines with the same color as the data points depict exponential fits.
        Plots (a) and (b) are for $P_{\text{exc}} = 1.7~P_\mathrm{thr}$ above threshold.
        The width $w$ of the selected phase-space region is $0.57$.
        Plots (c) and (d) are for $P_{\text{exc}} = 0.8~P_\mathrm{thr}$ below threshold.
        The width $w$ is $0.1$, except for the two highest radii $s$, where $w = 0.57$. 
        Insets in (a) and (c) show Husimi functions for those excitation powers from which the intensity $s$ is selected, cf. Fig. \ref{Fig:idea}.
        The inset in (d) displays a zoom of the region close to zero delay, revealing an oscillation of the mean amplitude.
        Right column exemplifies three $P_\Omega$ for $P_{\text{exc}} = 1.7~P_\mathrm{thr}$ and a selected radius $s = 9.9$ at time delays $\tau = -6~\mathrm{ps}$ (top), $\tau = 494~\mathrm{ps}$ (middle), and $\tau = 1\,214~\mathrm{ps}$ (bottom).
    }\label{Fig:delayplots}
\end{figure*}

\paragraph{Experimental time-resolved tomography.---}\hspace*{-2.5ex}
    Our system is a GaAs polariton microcavity \cite{MBADMSHS20, LPRSHSSA21,SM}.
    The sample is held in a cryostat at $10~\mathrm{K}$ and is excited nonresonantly at the first minimum of the stop band, with a linearly polarized continuous-wave laser.
    A nonresonant excitation prevents the system from inheriting coherence from the pump \cite{KMSL07,KRKBJKMSASSLDD06}.

    For our time-resolved tomography, the sample emission is divided into three homodyne detection channels, Fig. \ref{Fig:idea}.
    Two channels provide a proxy measurement of a Husimi function \cite{S01, LPRSHSSA21}, and the third one acts as a target arm \cite{TLA20}.
    Between the target and proxy, a temporal offset $\tau$ is controlled via a delay line.
    An annulus-shaped region of $Q(q_{ps},p_{ps})$, specified by radius $s$ and thickness $w$, selects specific intensities.
    For data in that region, we process the quadrature values $q$ from the target.
    Besides, we reconstruct the relative phase $\varphi$ between the local oscillator (LO) and signal in the target channel by adding the relative phase of the two selection channels, $\varphi_{ps} = \arctan(p_{ps}/q_{ps})$, to the relative phase $\Delta \varphi$ between the LOs in proxy and target channels.
    Then, the values $(q, \varphi)$ are fed into the previously described reconstruction scheme.
    This yields $P_\Omega$ as a function of the variable time delay $\tau$, providing information resolved for each selected intensity region beyond other approaches that average over intensities.

    In homodyne detection, one measures the overlap of signal and LO, resulting in a mode selectivity.
    This allows us to track the evolution of selected signal modes separately. 
    Here, our LO is aligned to the most dominant linear polarization mode in the emission.
    The LO's wavelength is resonant to the most intense zero-momentum ground-state mode of the polariton emission.
    The LO has a spectral full width at half maximum (FWHM) of $1.9~\mathrm{nm}$, a pulse duration of $460~\mathrm{fs}$ and a FWHM in $k$ space of $1.3~\mathrm{\mu m}^{-1}$, centered at $k = 0$. 

\paragraph*{Experimental results.---}\hspace*{-2.5ex}
    As an example, Fig. \ref{Fig:delayplots} (right column) shows reconstructed $P_\Omega$ functions for the highest excitation power $P_{\text{exc}} = 1.7~P_\mathrm{thr}$ \cite{comment} for three different time delays $\tau$.
    The phase diffuses with increasing $\tau$, resulting in the predicted broadening of the angular distribution.
    Simultaneously, the mean amplitude of $P_\Omega$ relaxes towards a steady-state value.

    To resolve the mean amplitude and the circular variance as functions of the selected intensity $s$ and time delays $\tau$, $\langle |\alpha| \rangle$ and $\mathrm{Var}(\phi)$ are directly derived from reconstructed $P_\Omega(\alpha)$ via Eq. \eqref{eq:cric_var} and $\langle |\alpha|\rangle=\int d^2\alpha\,P_\Omega(\alpha)|\alpha|$, using a discretized set of phase-space points $\alpha=q+ip$.
    The behavior of both quantities is plotted in Fig. \ref{Fig:delayplots}(a) and (b) for $P_{\text{exc}} = 1.7~P_\mathrm{thr}$, while $w=0.6$ was kept constant.
    Lines depict an exponential fit, which was found empirically to yield the best results compared to Gaussian and power-law fits \cite{SM}.
    $\mathrm{Var}(\phi)$ in plot (a) has its minimum around $\tau=0$, the smallest value being $0.14$ for the highest selected intensity, $s=11$.
    This minimum increases for smaller $s$ because of the fundamental phase-photon number uncertainty relation \cite{SM}.
    For increasing delays, the circular variance too increases but does not arrive at a uniform distribution (i.e., $\mathrm{Var}(\phi)=1$), even for the highest delay $\tau = 1\,214~\mathrm{ps}$.
    By contrast, the mean amplitude $\langle |\alpha| \rangle$ in plot (b) relaxes almost completely towards the steady-state value for large $\tau$, thus decaying even faster than the quantum phase.

    Figures \ref{Fig:delayplots}(c) and (d) show the results for a lower excitation power, $P_{\text{exc}} = 0.8~P_\mathrm{thr}$.
    For small $s$ values, the phase variance increases faster and reaches almost $1$, describing full phase decoherence.
    Meanwhile, the mean amplitude rapidly decays towards the steady-state value.
    Our method also reveals an oscillation around the stationary value, with a frequency of about $12.5~\mathrm{GHz}$.
    (This can be seen for $P_{\text{exc}} = P_\mathrm{thr}$ as well, but not for higher powers.)
    We attribute this effect to mode competition between modes of orthogonal linear polarizations \cite{SM}.
    A bistable regime of two cross-linear polarizations in a nonresonantly excited polariton condensate has also been proposed in Ref. \cite{S20}, leading to an oscillatory behavior of the condensate pseudospin components.
    But modulations of the spatial density of the polariton condensate, e.g., breathing modes, can also lie in this frequency range \cite{EPWSWPSTO21}.

\begin{figure}
    \includegraphics[width=0.5\textwidth]{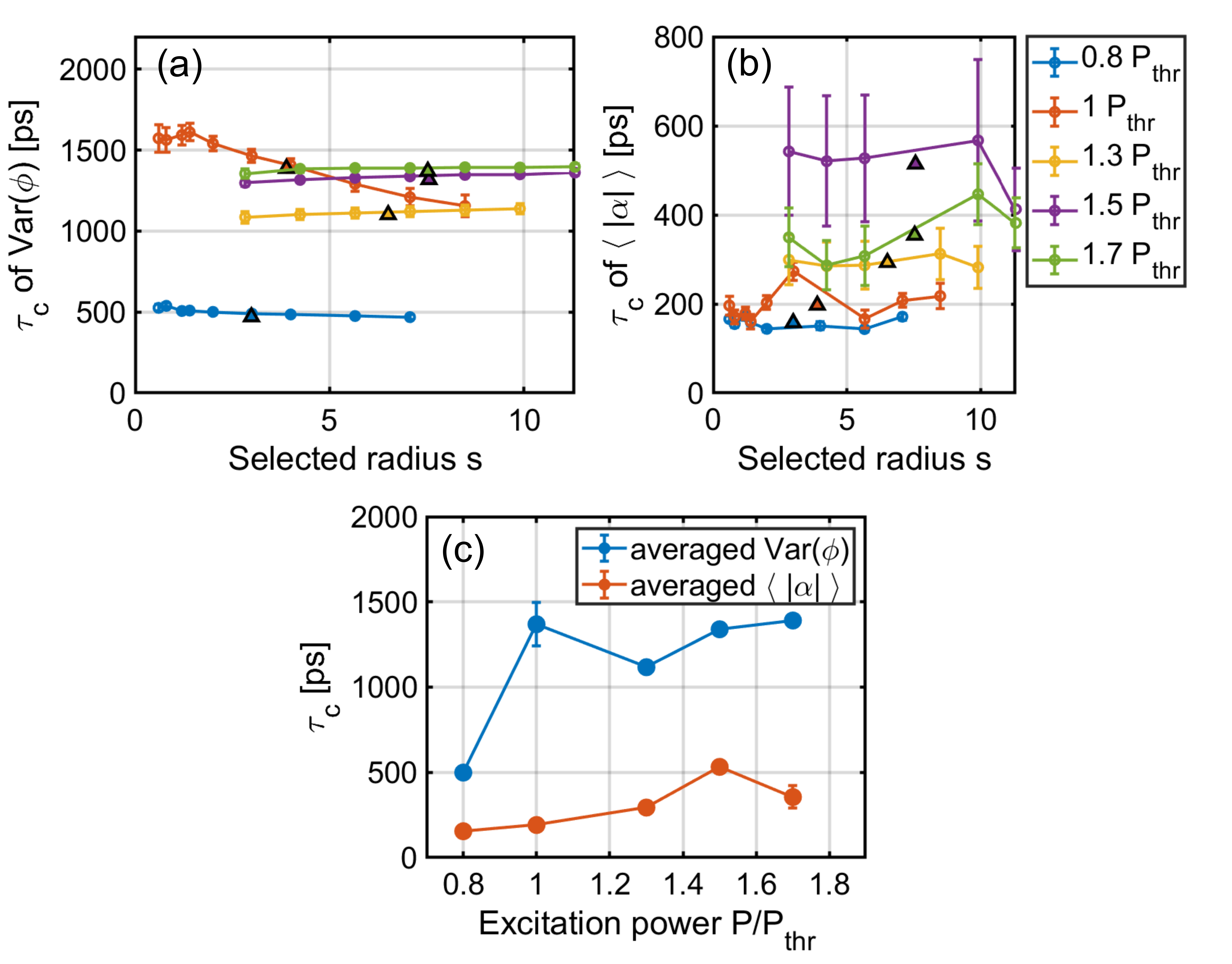}
    \caption{%
        (a) Decay times $\tau_c$ of $\mathrm{Var}(\phi)$ and (b) of the mean amplitude $\langle |\alpha| \rangle$ versus the selected radius $s$ for different excitation powers;
        connecting lines are guides for the eye.
        Triangles indicate the mean radius averaged over the Husimi distribution, i.e., the stationary state.
        For the amplitude decay times, $s$ values too close to the mean radius were discarded as fits are error-prone due to the flat curve shape in those cases.
        (c) Decay time $\tau_c$, averaged over the radii $s$, versus excitation power.
        For the average, the decay time for each radius has been weighted by the number of data points.
    }\label{fig:fitresults}
\end{figure}

    In addition, we analyze the temporal decay by fitting the temporal dependence of $\langle |\alpha| \rangle$ and $\mathrm{Var}(\phi)$ to exponential functions, $a \, \exp\left(-\tau/\tau_c\right) + d$, which estimates decay times $\tau_c$.
    In Figs. \ref{fig:fitresults}(a) and (b), this decay time is plotted versus the selected intensity radius $s$ for different excitation powers.
    For $\langle |\alpha| \rangle$, there is no significant dependence on $s$.
    In the phase variance, there is a slight tendency for states with higher $s$ to have a larger decay time for higher excitation powers, whereas the opposite trend appears for lower powers.
    This might be explained with a lower stability of the system for the lower powers and thus faster relaxation upon perturbations.
    At $P_{\text{exc}} = P_\mathrm{thr}$, the phase decay time is exceptionally long.
    Here, at the transition between thermal and coherent state, the system is least affected by heating and interaction with the reservoir, as well as polariton-polariton interactions, which become relevant decoherence mechanisms at higher excitation powers.

    In Fig. \ref{fig:fitresults}(c), the decay times were averaged over $s$ and plotted versus the excitation power.
    Clearly, the phase and the amplitude diffuse with different speeds, the amplitude having the shorter decay time.
    Notably, the mean phase coherence time is exceptionally long, between $520~\mathrm{ps}$ and $1\,390~\mathrm{ps}$.
    In earlier studies, coherence times on the order of $100~\mathrm{ps}$ were reported by employing a noise-free single-mode excitation laser \cite{LKWBSRBSWAD08} and a spatially confined cavity that favors single-mode emission \cite{KZWFBKSHD16}.
    By condensate trapping with a patterned pump, diminishing interactions with the reservoir, nanosecond coherence times were demonstrated \cite{APAZLLL19,OTPSO21,BZSGTAL22}.
    Top-hat-shaped pump lasers led to coherence times of up to $90~\mathrm{ps}$, causing less spread of the condensate in $k$ space and longer coherence times \cite{APAZLLL19}.
    However, most of these studies, except Ref. \cite{OTPSO21}, reported a Gaussian shape of $g^{(1)}(\tau)$ \cite{BDKDKOFSSH20}.
    And, in Ref. \cite{KZWFBKSHD16}, the observed decay is exponential above threshold but Gaussian for higher excitation powers.
    This Gaussian decay suggests inhomogeneous signal broadening.
    Compared to these results, we attribute our long coherence times to a combination of using a single-mode excitation laser with a relatively large beam diameter, leading to a small spread in $k$ space and less reservoir density, as well as filtering our signal with the LO to remove inhomogeneous broadening.

\begin{figure}
    \includegraphics[width=\columnwidth]{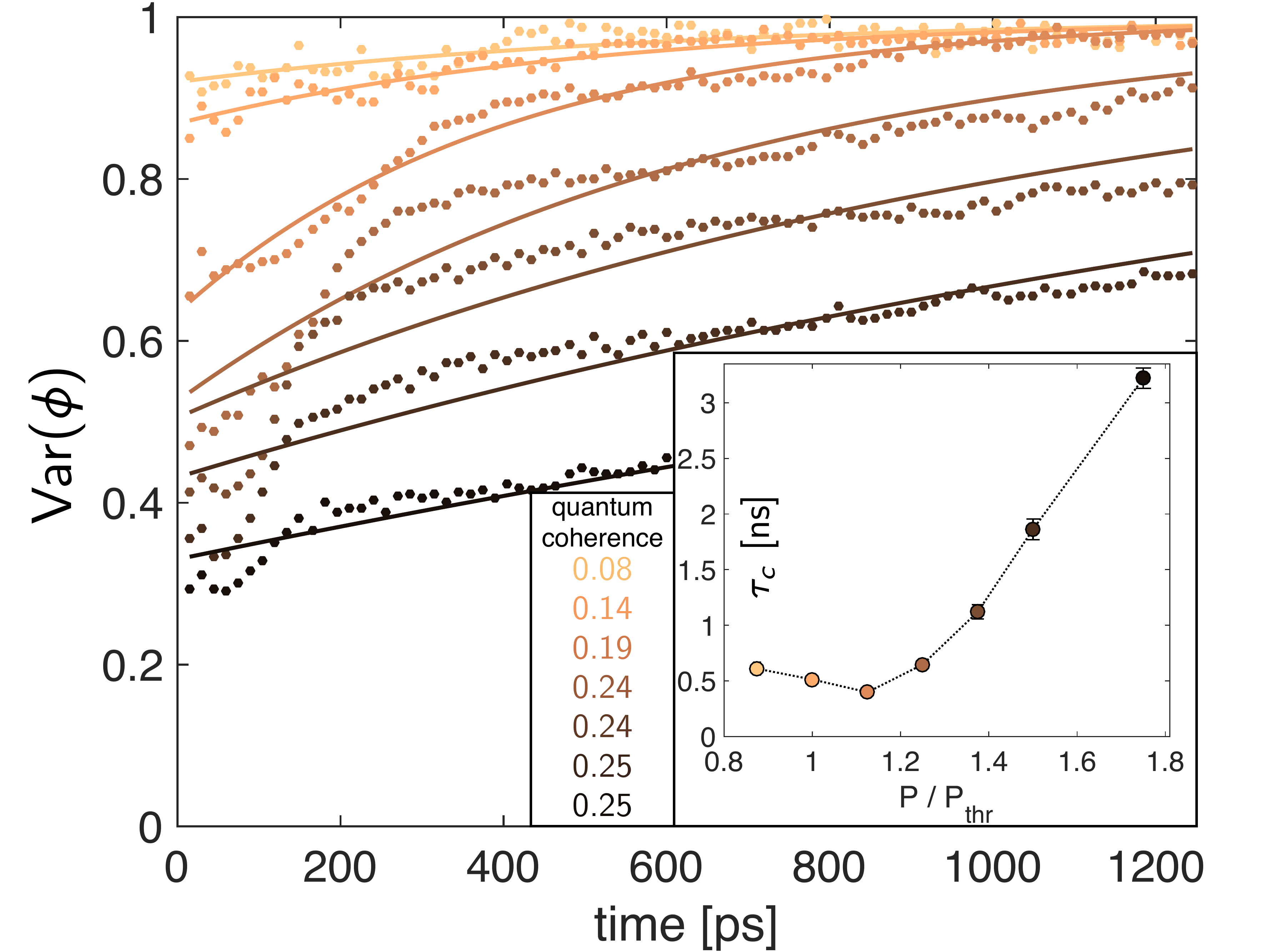}
    \caption{%
        Numerical results. 
        Circular variance $\mathrm{Var}(\phi)$ from $P_\Omega$ for the $k=0$ condensate mode as a function of time for different excitation powers.
        Lines are the fitted exponential curves, color-coded to the excitation powers as defined in the inset;
        the inset shows the fitted coherence times $\tau_c$ as a function of the excitation power, with error margins of one standard deviation.
        Additionally, the values of quantum coherence are given.
    }\label{fig:Num}
\end{figure}

\paragraph{Simulation.---}\hspace*{-2.5ex}
    We also carried out theoretical simulations utilizing a stochastic Gross--Pitaevskii model, based on the truncated Wigner approximation \cite{SLC02,BBDBG08}.
    This phase-space method was successfully employed to study coherence properties of polariton condensates \cite{CC05,WS09,CDZCPS18,LPRSHSSA21}.
    Similar to the experiment, the phase variance is dynamically tracked to probe the decay of quantum coherence.
    We initialize the system as a displaced thermal state that then evolves in time.
    The mean displacement and the standard deviation of the initial distribution are determined through the steady-state values for the mean polariton number and the quantum coherence \cite{SM}.

    Importantly, we employ a convolution-deconvolution approach to derive a relation between our $P_\Omega$ function and the Wigner function that unifies the existing simulations with our method \cite{SM}.
    While the deconvolution part from the Wigner function to the $P$ distribution is not well-behaved because of diverging integrals, we find that the following convolution with our non-Gaussian $\Omega$ in Eq. \eqref{eq:ChosenKernel} eventually yields well-defined expressions, being paramount for numerical analyses.

    Figure \ref{fig:Num} shows the numerical results for the time-dependent circular variance and the correspondingly fitted coherence time $\tau_c$ of the zero-momentum condensate mode for different excitation powers.
    $\tau_c$ increases significantly around the threshold excitation power $P_{\mathrm{thr}}$ and reaches values from $0.5~\mathrm{ns}$ to $1.5~\mathrm{ns}$, being comparable to the experiment until about $1.5~P_{\mathrm{thr}}$.
    The source of decoherence can be attributed to the nonlinear part of the effective potential, caused by interactions of the polaritons with themselves and the reservoir.
    Density fluctuations, in turn, lead to frequency fluctuations in the condensate mode that constitutes an intrinsic decoherence mechanism \cite{LKWBSRBSWAD08}, which is particularly relevant for lower polariton densities near the threshold.
    Unlike the experiment, however, the coherence time in the simulation does not saturate but continues to increase beyond $1.5~P_{\mathrm{thr}}$.
    This is caused by the lack of decoherence mechanisms in the numerical model which become experimentally important at higher excitation powers, e.g., temperature effects due to sample heating and higher-order scattering processes, as discussed above.

    We further find that $\tau_c$ decreases with increasing interaction strength $g_c$ and with decreasing condensate-reservoir interaction $g_r$.
    These trends are an effect of the effective potential which causes the decoherence in the numerical model. 
    Up to first order, the nonlinear part of the effective potential reads $V_{\mathrm{nl}}= g_c[1-P_{\mathrm{pump}}^*(g_r\gamma_c)/(g_c\gamma_r)]|\psi|^2$ \cite{DFDLBKSHO14}, with $P_{\mathrm{pump}}^{*}\equiv P_{\mathrm{pump}}R_r /(\gamma_c\gamma_r)$, revealing the observed parameter trends.
    The coherence time also depends on the spatial pump profile, altering the spectral shape of the emission \cite{WCC08}.
    Moreover, we find that, in the presence of a realistic static disorder potential, the condensate emission is slightly shifted to finite $k$ values, but the coherence times of the maximum-intensity mode remain similar to the values without disorder.

\paragraph{Conclusion.---}\hspace*{-2.5ex}
    By combining quantum dynamics, modern phase-space methods, and coherence quantifiers of quantum-technological resourcefulness, we explored the time evolution of polariton condensates.
    We implemented a phase-space-based approach to access quantum coherence via circular variances and to characterize the state in general, while also overcoming limitations of other phase-space reconstruction techniques.
    And we devised a numerical approach that renders state-of-the-art simulations compatible to our methodology.
    Thereby, an advanced and complete framework to dynamically track the quantum-informational resourcefulness of polariton condensates is established and implemented.

    Experimentally, we employed multi-channel homodyne detection to directly reconstruct advanced phase-space representations.
    While homodyne detection allows one to selectively filter on one mode for measurement, our conditional spectroscopy further enabled us to analyze the dynamics of the state as a function of the initial intensity.
    We thereby observed long decay times of quantum coherence on the order of $1\,\mathrm{ns}$, agreeing well with our numerical simulations.

    Combining the here-developed strategies with our multi-channel detection \cite{TLA20} and generalized multi-mode phase-space methods \cite{ASV13} further paves the way for future investigations, such as the interplay between different condensate modes to explore mode competition \cite{BZSGTAL22} and the dynamical characterization of multimode quantum correlations \cite{KASSVSH21}.
    Furthermore, our approach can also advance the study of other quantum systems beyond the specific scenario explored here, such as trapped ions \cite{FH20} and ensembles of atoms \cite{KVCBAP12}.

\paragraph{Acknowledgments.---}\hspace*{-2.5ex}
	The authors acknowledge funding through the Deutsche Forschungsgemeinschaft (DFG, German Research Foundation) via the transregional collaborative research center TRR~142 (Projects A04 and C10, Grant No. 231447078).
	A grant for computing time at the Paderborn Center for Parallel Computing (PC\textsuperscript{2}) is gratefully acknowledged.


\begin{thebibliography}{99}

	\bibitem{SAP17}
		A. Streltsov, G. Adesso, and M. B. Plenio,
		Colloquium: Quantum coherence as a resource,  
		\href{https://doi.org/10.1103/RevModPhys.89.041003}{Rev. Mod. Phys. \textbf{89}, 041003 (2017)}.
	\bibitem{CG19}
		E. Chitambar and G. Gour,
		Quantum resource theories,
		\href{https://doi.org/10.1103/RevModPhys.91.025001}{Rev. Mod. Phys. \textbf{91}, 025001 (2019)}.
	\bibitem{BCP14}
		T. Baumgratz, M. Cramer, and M. B. Plenio,
		Quantifying Coherence,
		\href{https://doi.org/10.1103/PhysRevLett.113.140401}{Phys. Rev. Lett. \textbf{113}, 140401 (2014)}.
	\bibitem{LM14}
		F. Levi and F. Mintert,
		A quantitative theory of coherent delocalization,
		\href{https://doi.org/10.1088/1367-2630/16/3/033007}{New J. Phys. \textbf{16}, 033007 (2014)}.
	\bibitem{SV15}
		J. Sperling and W. Vogel,
		Convex ordering and quantification of quantumness,
		\href{https://doi.org/10.1088/0031-8949/90/7/07402}{Phys. Scr. \textbf{90}, 074024 (2015)}.
	\bibitem{WY16}
		A. Winter and D. Yang,
		Operational Resource Theory of Coherence,
		\href{https://doi.org/10.1103/PhysRevLett.116.120404}{Phys. Rev. Lett. \textbf{116}, 120404 (2016)}.
	\bibitem{NC00}
		M. A. Nielsen and I. L. Chuang,
		\textit{Quantum Computation and Quantum Information}, 10th ed.
		(\href{https://doi.org/10.1017/CBO9780511976667}{Cambridge University Press, Cambridge, England, 2010}).
	\bibitem{DM03}
		J. P. Dowling and J. G. Milburn,
		Quantum technology: the second quantum revolution,
		\href{https://doi.org/10.1098/rsta.2003.1227}{Phil. Trans. R. Soc. Lond. A \textbf{361}, 1655 (2003)}.
	\bibitem{D20}
		I. H. Deutsch,
		Harnessing the Power of the Second Quantum Revolution,
		\href{https://doi.org/10.1103/PRXQuantum.1.020101}{PRX Quantum \textbf{1}, 020101 (2020)}.
	\bibitem{BBCJPW93}
		C. H. Bennett, G. Brassard, C. Cr\'epeau, R. Jozsa, A. Peres, and W. K. Wootters,
		Teleporting an unknown quantum state via dual classical and Einstein-Podolsky-Rosen channels,
		\href{https://doi.org/10.1103/PhysRevLett.70.1895}{Phys. Rev. Lett. \textbf{70}, 1895 (1993)}.
	\bibitem{S94}
		P. W. Shor,
		Polynomial-Time Algorithms for Prime Factorization and Discrete Logarithms on a Quantum Computer,
		\href{https://doi.org/10.1137/S0097539795293172}{SIAM J. Sci. Statist. Comput. \textbf{26}, 1484 (1997)}.
	\bibitem{BB84}
		C. H. Bennett and G. Brassard,
		Quantum cryptography: Public key distribution and coin tossing,
		\href{https://doi.org/10.1016/j.tcs.2014.05.025}{Theor. Comput. Sci. \textbf{560} (Part 1), 7 (2014)}.
	\bibitem{LPRSHSSA21}
		C. L\"uders, M. Pukrop, E. Rozas, C. Schneider, S. H\"ofling, J. Sperling, S. Schumacher, and M. A\ss{}mann,
		Quantifying quantum coherence in polariton condensates,
		\href{https://doi.org/10.1103/PRXQuantum.2.030320}{PRX Quantum \textbf{2}, 030320 (2021)}.
	\bibitem{SV20}
		J. Sperling and W. Vogel,
		Quasiprobability distributions for quantum-optical coherence and beyond,
		\href{https://doi.org/10.1088/1402-4896/ab5501}{Phys. Scr. \textbf{95}, 034007 (2020)}.
	\bibitem{SV05}
	    E. V. Shchukin and W. Vogel
	    Nonclassical moments and their measurement,
        \href{https://doi.org/10.1103/PhysRevA.72.043808}{Phys. Rev. A \textbf{72}, 043808 (2005)}.
	\bibitem{CC13}
		I. Carusotto and C. Ciuti,
		Quantum fluids of light,
		\href{https://doi.org/10.1103/RevModPhys.85.299}{Rev. Mod. Phys. \textbf{85}, 299 (2013)}.
	\bibitem{DHY10}
		H. Deng, H. Haug, and Y. Yamamoto,
		Exciton-polariton Bose-Einstein condensation,
		\href{https://doi.org/10.1103/RevModPhys.82.1489}{Rev. Mod. Phys. \textbf{82}, 1489 (2010)}.
	\bibitem{APAZLLL19}
		A. Askitopoulos, L. Pickup, S. Alyatkin, A. Zasedatelev, K. G. Lagoudakis, W. Langbein, and P. G. Lagoudakis,
		Giant increase of temporal coherence in optically trapped polariton condensate,
		\href{https://doi.org/10.48550/arXiv.1911.08981}{arXiv:1911.08981}.
	\bibitem{LKWBSRBSWAD08}
		A. P. D. Love, D. N. Krizhanovskii, D. M. Whittaker, R. Bouchekioua, D. Sanvitto, S. Al Rizeiqi, R. Bradley, M. S. Skolnick, P. R. Eastham, R. Andr\'e, and L. S. Dang,
		Intrinsic Decoherence Mechanisms in the Microcavity Polariton Condensate,
		\href{https://doi.org/10.1103/PhysRevLett.101.067404}{Phys. Rev. Lett. \textbf{101}, 067404 (2008)}.
	\bibitem{OTPSO21}
		K. Orfanakis, A. F. Tzortzakakis, D. Petrosyan, P. G. Savvidis, and H. Ohadi,
		Ultralong temporal coherence in optically trapped exciton-polariton condensates,
		\href{https://doi.org/10.1103/PhysRevB.103.235313}{Phys. Rev. B \textbf{103}, 235313 (2021)}.
	\bibitem{KZWFBKSHD16}
		S. Kim, B. Zhang, Z. Wang, J. Fischer, S. Brodbeck, M. Kamp, C. Schneider, S. H\"ofling, and H. Deng,
		Coherent Polariton Laser,
		\href{https://doi.org/10.1103/PhysRevX.6.011026}{Phys. Rev. X \textbf{6}, 011026 (2016)}.
	\bibitem{KFASBAKSH18}
	    M. Klaas, H. Flayac, M. Amthor, I. G. Savenko, S. Brodbeck, T. Ala-Nissila, S. Klembt, C. Schneider, and S. H\"ofling,
        Evolution of Temporal Coherence in Confined Exciton-Polariton Condensates,
        \href{https://doi.org/10.1103/PhysRevLett.120.017401}{Phys. Rev. Lett. \textbf{120}, 017401 (2018)}.
	\bibitem{BZSGTAL22}
		S. Baryshev, A. Zasedatelev, H. Sigurdsson, I. Gnusov, J. D. T\"opfer, A. Askitopoulos, and P.  G. Lagoudakis,
		Engineering Photon Statistics in a Spinor Polariton Condensate,
		\href{https://doi.org/10.1103/PhysRevLett.128.087402}{Phys. Rev. Lett. \textbf{128}, 087402 (2022)}.
	\bibitem{SLC02}
		A. Sinatra, C. Lobo, and Y. Castin,
		The truncated Wigner method for Bose-condensed gases: limits of validity and applications,
		\href{https://doi.org/10.1088/0953-4075/35/17/301}{J. Phys. B: At. Mol. Opt. Phys. \textbf{35}, 3599 (2002)}.
	\bibitem{BBDBG08}
		P. B. Blakie, A. S. Bradley, M. J. Davis, R. J. Ballagh, and C. W. Gardiner,
		Dynamics and statistical mechanics of ultra-cold Bose gases using c-field techniques,
		\href{https://doi.org/10.1080/00018730802564254}{Adv. Phys. \textbf{57}, 363 (2008)}.
	\bibitem{G63}
		R. J. Glauber,
		Coherent and Incoherent States of the Radiation Field,
		\href{https://doi.org/10.1103/PhysRev.131.2766}{Phys. Rev. \textbf{131}, 2766 (1963)}.
	\bibitem{S63}
		E. C. G. Sudarshan,
		Equivalence of Semiclassical and Quantum Mechanical Descriptions of Statistical Light Beams,
		\href{https://doi.org/10.1103/PhysRevLett.10.277}{Phys. Rev. Lett. \textbf{10}, 277 (1963)}.
	\bibitem{S16}
		J. Sperling,
		Characterizing maximally singular phase-space distributions,
		\href{https://doi.org/10.1103/PhysRevA.94.013814}{Phys. Rev. A \textbf{94}, 013814 (2016)}.
	\bibitem{CG69}
		K. E. Cahill and R. J. Glauber,
		Density Operators and Quasiprobability Distributions,
		\href{https://doi.org/10.1103/PhysRev.177.1882}{Phys. Rev. \textbf{177}, 1882 (1969)}.
	\bibitem{AW70}
		G. S. Agarwal and E. Wolf,
		Calculus for Functions of Noncommuting Operators and General Phase-Space Methods in Quantum Mechanics. II. Quantum Mechanics in Phase Space,
		\href{https://doi.org/10.1103/PhysRevD.2.2187}{Phys. Rev. D \textbf{2}, 2187 (1970)}.
	\bibitem{T97}
		 S. M. Tan,
		 An inverse problem approach to optical homodyne tomography,
		 \href{https://doi.org/10.1080/09500349708231881}{J. Mod. Opt. \textbf{44}, 2233 (1997)}.
	\bibitem{SSG09}
		V. N. Starkov, A. A. Semenov, and H. V. Gomonay,
		Numerical reconstruction of photon-number statistics from photocounting statistics: Regularization of an ill-posed problem,
		\href{https://doi.org/10.1103/PhysRevA.80.013813}{Phys. Rev. A \textbf{80}, 013813 (2009)}.
	\bibitem{R96}
		T. Richter,
		Pattern functions used in tomographic reconstruction of photon statistics revisited,
		\href{https://doi.org/10.1016/0375-9601(96)00029-1}{Phys. Lett. A \textbf{211}, 327 (1996)}.
	\bibitem{LMKRR96}
		U. Leonhard, M. Munroe, T. Kiss, T. Richter, and M. G. Raymer,
		Sampling of photon statistics and density matrix using homodyne detection,
		\href{https://doi.org/10.1016/0030-4018(96)00061-2}{Opt. Commun. \textbf{127}, 144 (1996)}.
	\bibitem{H97}
		Z. Hradil,
		Quantum-state estimation,
		\href{https://doi.org/10.1103/PhysRevA.55.R1561}{Phys. Rev. A \textbf{55}, R1561(R) (1997)}.
	\bibitem{L04}
		A. I. Lvovsky,
		Iterative maximum-likelihood reconstruction in quantum homodyne tomography,
		\href{https://doi.org/10.1088/1464-4266/6/6/014}{J. Opt. B \textbf{6}, S556 (2004)}.
	\bibitem{KWR04}
		R. Kosut, I. A. Walmsley, and H. Rabitz,
		Optimal Experiment Design for Quantum State and Process Tomography and Hamiltonian Parameter Estimation,
		\href{https://arxiv.org/abs/quant-ph/0411093}{arXiv:0411093 [quant-ph]}.
	\bibitem{KV10}
		T. Kiesel and W. Vogel,
		Nonclassicality filters and quasiprobabilities,
		\href{https://doi.org/10.1103/PhysRevA.82.032107}{Phys. Rev. A \textbf{82}, 032107 (2010)}.
	\bibitem{KVTMS21}
		B. K\"uhn, W. Vogel, V. Thiel, S. Merkouche, and B. J. Smith,
		Gaussian versus Non-Gaussian Filtering of Phase-Insensitive Nonclassicality,
		\href{https://doi.org/10.1103/PhysRevLett.126.173603}{Phys. Rev. Lett. \textbf{126}, 173603 (2021)}.
	\bibitem{KASSVSH21}
		S. K\"ohnke, E. Agudelo, M. Sch\"unemann, O. Schlettwein, W. Vogel, J. Sperling, and B. Hage,
		Quantum Correlations beyond Entanglement and Discord,
		\href{https://doi.org/10.1103/PhysRevLett.126.170404}{Phys. Rev. Lett. \textbf{126}, 170404 (2021)}.
	\bibitem{KK08}
	    M. Kira and S. W. Koch,
	    Cluster-Expansion Representation in Quantum Optics,
	    \href{https://link.aps.org/doi/10.1103/PhysRevA.78.022102}{Phys. Rev. A \textbf{78}, 022102 (2008)}.
	\bibitem{KKS11}
	    M. Kira, S. W. Koch, R. P. Smith, A. E. Hunter, and S. T. Cundiff,
	    Quantum Spectroscopy with Schr\"odinger Cat States,
	    \href{https://doi.org/10.1038/nphys2091}{Nat. Phys. \textbf{7}, 799 (2011)}.
	\bibitem{HLC14}
	    A. E. Almand-Hunter, H. Li, S. T. Cundiff, M. Mootz, S. Kira, and S. W. Koch,
	    Quantum Droplets of Electrons and Holes,
	    \href{https://doi.org/10.1038/nature12994}{Nature (London) \textbf{506}, 471 (2014)}.
	\bibitem{KVHS11}
		T. Kiesel, W. Vogel, B. Hage, and R. Schnabel,
		Direct Sampling of Negative Quasiprobabilities of a Squeezed State,
		\href{https://doi.org/10.1103/PhysRevLett.107.113604}{Phys. Rev. Lett. \textbf{107}, 113604 (2011)}.
	\bibitem{SM} 
		Supplemental Material, which includes Refs. \cite{KV10,VW06,KVHS11,KV12a,KV12b,ASVKMH15,LPRSHSSA21,SLC02,BBDBG08,WS09,CC13,MBADMSHS20,HP06,KRKBJKMSASSLDD06,WCC08,TLA20,KBMCHL12,S01,SKL07,CBDSGDWPGLSS18,SBCR93}, for additional technical considerations, pertaining to the method of regularized phase-space functions (incl. reconstruction techniques and relations to quantum coherence), our generalized numeric toolbox, as well as experimental details.
	\bibitem{F93}
		N. I. Fisher,
		\textit{Statistical Analysis of Circular Data}
		(\href{https://doi.org/10.1017/CBO9780511564345}{Cambridge University Press, Cambridge, UK, 1993}).
	\bibitem{MBADMSHS20}
		X. Ma, B. Berger, M. A\ss{}mann, R. Driben, T. Meier, C. Schneider, S. H\"ofling, and S. Schumacher,
		Realization of all-optical vortex switching in exciton-polariton condensates,
		\href{https://doi.org/10.1038/s41467-020-14702-5}{Nat. Commun. \textbf{11}, 897 (2020)}. 
	\bibitem{KMSL07}
		J. Keeling, F. M. Marchetti, M. H. Szyma\'{n}ska, and P. B. Littlewood,
		Collective coherence in planar semiconductor microcavities,
		\href{https://doi.org/10.1088/0268-1242/22/5/R01}{Semicond. Sci. Technol. \textbf{22}, R1 (2007)}.
	\bibitem{KRKBJKMSASSLDD06}
		J. Kasprzak, M. Richard, S. Kundermann, A. Baas, P. Jeambrun, J. M. J. Keeling, F. M. Marchetti, M. H. Szym{\'{a}}nska, R. Andr{\'{e}}, J. L. Staehli, V. Savona, P. B. Littlewood, B. Deveaud, and L. S. Dang,
		Bose–Einstein condensation of exciton polaritons,
		\href{https://doi.org/10.1038/nature05131}{Nature (London) \textbf{443}, 409 (2006)}.
	\bibitem{S01}
		W. P. Schleich,
		\textit{Quantum Optics in Phase Space}
		(\href{https://doi.org/10.1002/3527602976}{John Wiley \& Sons, Berlin, 2001}).
	\bibitem{TLA20}
		J. Thewes, C. L\"uders, and M. A\ss{}mann,
		Conditional spectroscopy via nonstationary optical homodyne quantum state tomography,
		\href{https://doi.org/10.1103/PhysRevA.101.023824}{Phys. Rev. A \textbf{101}, 023824 (2020)}.
    \bibitem{comment}
        Note that, for our purpose, we define the threshold power not via a nonlinear increase of the total emission but via the onset of quantum coherence in the specific mode filtered by the LO (cf. \cite{SM} for details).
        Thus, the threshold is meaningful for the mode that we actually measure and directly comparable to our simulation, being based on the same definition of $P_{\mathrm{thr}}$.
	\bibitem{S20}
		H. Sigurdsson,
		Hysteresis in linearly polarized nonresonantly driven exciton-polariton condensates,
		\href{https://doi.org/10.1103/PhysRevResearch.2.023323}{Phys. Rev. Research \textbf{2}, 023323 (2020)}.
	\bibitem{EPWSWPSTO21}
		E. Estrecho, M. Pieczarka, M. Wurdack, M. Steger, K. West, L. N. Pfeiffer, D. W. Snoke, A. G. Truscott, and E. A. Ostrovskaya,
		Low-Energy Collective Oscillations and Bogoliubov Sound in an Exciton-Polariton Condensate,
		\href{https://doi.org/10.1103/PhysRevLett.126.075301}{Phys. Rev. Lett. \textbf{126}, 075301 (2021)}.
    \bibitem{BDKDKOFSSH20}
	    S. Betzold, M. Dusel, O. Kyriienko, C. P. Dietrich, S. Klembt, J. Ohmer, U. Fischer, I. A. Shelykh, C. Schneider, and S. H\"ofling,
	    Coherence and Interaction in Confined Room-Temperature Polariton Condensates with Frenkel Excitons,
	    \href{https://doi.org/10.1021/acsphotonics.9b01300}{ACS Photonics \textbf{7}, 284 (2020)}.
	\bibitem{WS09}
		M. Wouters and V. Savona,
		Stochastic classical field model for polariton condensates,
		\href{https://doi.org/10.1103/PhysRevB.79.165302}{Phys. Rev. B \textbf{79}, 165302 (2009)}.
	\bibitem{CC05}
		I. Carusotto and C. Ciuti,
		Spontaneous microcavity-polariton coherence across the parametric threshold: Quantum Monte Carlo studies,
		\href{https://doi.org/10.1103/PhysRevB.72.125335}{Phys. Rev. B \textbf{72}, 125335 (2005)}.
	\bibitem{CDZCPS18}
		P. Comaron, G. Dagvadorj, A. Zamora, I. Carusotto, N. P. Proukakis, and M. H. Szyma\'nska,
		Dynamical Critical Exponents in Driven-Dissipative Quantum Systems,
		\href{https://doi.org/10.1103/PhysRevLett.121.095302}{Phys. Rev. Lett. \textbf{121}, 095302 (2018)}.
	\bibitem{DFDLBKSHO14}
		R. Dall, M. D. Fraser, A. S. Desyatnikov, G. Li, S. Brodbeck, M. Kamp, C. Schneider, S. H\"ofling, and E. A. Ostrovskaya,
		Creation of Orbital Angular Momentum States with Chiral Polaritonic Lenses,
		\href{https://doi.org/10.1103/PhysRevLett.113.200404}{Phys. Rev. Lett. \textbf{113}, 200404 (2014)}.
	\bibitem{WCC08}
		M. Wouters, I. Carusotto, and C. Ciuti,
		Spatial and spectral shape of inhomogeneous nonequilibrium exciton-polariton condensates,
		\href{https://doi.org/10.1103/PhysRevB.77.115340}{Phys. Rev. B \textbf{77}, 115340 (2008)}.
	\bibitem{ASV13}
		E. Agudelo, J. Sperling, and W. Vogel,
		Quasiprobabilities for multipartite quantum correlations of light,
		\href{https://doi.org/10.1103/PhysRevA.87.033811}{Phys. Rev. A \textbf{87}, 033811 (2013)}.
    \bibitem{FH20}
        C. Fl\"uhmann and J. P. Home,
        Direct Characteristic-Function Tomography of Quantum States of the Trapped-Ion Motional Oscillator,
        \href{https://doi.org/10.1103/PhysRevLett.125.043602}{Phys. Rev. Lett. \textbf{125}, 043602 (2020)}.
    \bibitem{KVCBAP12}
        T. Kiesel, W. Vogel, S. L. Christensen, J.-B. B\'eguin, J. Appel, and E. S. Polzik,
        Atomic nonclassicality quasiprobabilities,
        \href{https://doi.org/10.1103/PhysRevA.86.042108}{Phys. Rev. A \textbf{86}, 042108 (2012)}.
	\bibitem{VW06}
		W. Vogel and D.-G. Welsch,
		\textit{Quantum Optics}, 3rd ed.
		(\href{https://doi.org/10.1002/3527608524}{Wiley‐VCH Verlag, Weinheim, 2006}).
	\bibitem{KV12a}
		T. Kiesel and W. Vogel,
		Universal nonclassicality witnesses for harmonic oscillators,
		\href{https://doi.org/10.1103/PhysRevA.85.062106}{Phys. Rev. A \textbf{85}, 062106 (2012)}.
	\bibitem{KV12b}
		T. Kiesel and W. Vogel,
		Complete nonclassicality test with a photon-number-resolving detector,
		\href{https://doi.org/10.1103/PhysRevA.86.032119}{Phys. Rev. A \textbf{86}, 032119 (2012)}.
	\bibitem{ASVKMH15}
		E. Agudelo, J. Sperling, W. Vogel, S. K\"ohnke, M. Mraz, and B. Hage,
		Continuous sampling of the squeezed-state nonclassicality,
		\href{https://doi.org/10.1103/PhysRevA.92.033837}{Phys. Rev. A \textbf{92}, 033837 (2015)}.
	\bibitem{HP06}
		J. A. Hansen and C. Penland,
		Efficient Approximate Techniques for Integrating Stochastic Differential Equations,
	    \href{https://doi.org/10.1175/MWR3192.1}{Mon. Weather Rev. \textbf{134}, 3006 (2006)}.
	\bibitem{KBMCHL12}
		R. Kumar, E. Barrios, A. MacRae, E. Cairns, E. H. Huntington, and A. I. Lvovsky,
		Versatile wideband balanced detector for quantum optical homodyne tomography,
		\href{https://doi.org/10.1016/j.optcom.2012.07.103}{Opt. Commun. \textbf{285}, 5259 (2012)}.
	\bibitem{SKL07}
		M. H. Szyma\'{n}ska, J. Keeling, and P. B. Littlewood,
		Mean-field theory and fluctuation spectrum of a pumped decaying Bose-Fermi system across the quantum condensation transition,
		\href{https://doi.org/10.1103/PhysRevB.75.195331}{Phys. Rev. B \textbf{75}, 195331 (2007)}.
	\bibitem{CBDSGDWPGLSS18}
		D. Caputo, D. Ballarini, G. Dagvadorj, C. S\'{a}nchez Mu\~{n}oz, M. De Giorgi, L. Dominici, K. West, L. N. Pfeiffer, G. Gigli, F. P. Laussy, M. H. Szyma\'{n}ska, and D. Sanvitto,
		Topological order and thermal equilibrium in polariton condensates,
		\href{https://doi.org/10.1038/nmat5039}{Nat. Mater. \textbf{17}, 145 (2018)}.
	\bibitem{SBCR93}
		D. T. Smithey, M. Beck, J. Cooper, and M. G. Raymer,
		Measurement of number-phase uncertainty relations of optical fields,
		\href{https://doi.org/10.1103/PhysRevA.48.3159}{Phys. Rev. A \textbf{48}, 3159 (1993)}.	

\end{thebibliography}
\end{document}


\title{Supplemental Material}

\author{Carolin L\"uders}
	\affiliation{Experimentelle Physik 2, Technische Universit\"at Dortmund, D-44221 Dortmund, Germany}
\author{Matthias Pukrop}
	\affiliation{Department of Physics and Center for Optoelectronics and Photonics Paderborn (CeOPP), Universit\"at Paderborn, 33098 Paderborn, Germany}
\author{Franziska Barkhausen}
	\affiliation{Department of Physics and Center for Optoelectronics and Photonics Paderborn (CeOPP), Universit\"at Paderborn, 33098 Paderborn, Germany}
\author{Elena Rozas}
	\affiliation{Experimentelle Physik 2, Technische Universit\"at Dortmund, D-44221 Dortmund, Germany}
\author{Christian Schneider}
    \affiliation{Institute of Physics, University of Oldenburg, D-26129 Oldenburg, Germany}
\author{Sven H\"ofling}
    \affiliation{Technische Physik, Physikalisches Institut and W\"urzburg-Dresden Cluster of Excellence ct.qmat, Universit\"at W\"urzburg, 97074 W\"urzburg, Germany}
\author{Jan Sperling}
    \affiliation{Theoretical Quantum Science, Institute for Photonic Quantum Systems (PhoQS), Paderborn University, Warburger Stra\ss{}e 100, 33098 Paderborn, Germany}
\author{Stefan Schumacher}
	\affiliation{Department of Physics and Center for Optoelectronics and Photonics Paderborn (CeOPP), Universit\"at Paderborn, 33098 Paderborn, Germany}
	\affiliation{Wyant College of Optical Sciences, University of Arizona, Tucson, Arizona 85721, USA}
\author{Marc A\ss{}mann}
	\affiliation{Experimentelle Physik 2, Technische Universit\"at Dortmund, D-44221 Dortmund, Germany}

\begin{abstract}
	Here, we provide additional technical details to complement the key findings as reported on in the main text.
	This supplemental document is structured as follows:
	In Sec. \ref{supp.sec:regP}, a brief introduction to the method of regularized $P$ functions is provided, together with its specific properties that are essential in the context of our study.
	Section \ref{supp.sec:numerics} generalizes the numeric toolbox that is commonly used to evolve the Wigner function in time, rendering this approach applicable to regularized $P$ functions as well.
	In Sec. \ref{supp.sec:Implementation}, we provide additional experimental details.
\end{abstract}

\date{\today}
\maketitle
\appendix


\section{Regularized $P$ functions}\label{supp.sec:regP}

	In this section, we provide technical details on the assessment of quantum coherence in terms of regularized phase-space functions and their reconstructions from data.

\subsection{Coherence properties under convolutions}

	In this first step, we demonstrate that our regularization of the Glauber-Sudarshan distribution is consistent with the assessment of quantum coherence, constituting the key feature of our approach.

	It is relatively straightforward to show that a state is an incoherent mixture of photon number states, $\hat\rho=\sum_{n\in\mathbb N} p_n|n\rangle\langle n|$, if and only if the corresponding Glauber-Sudarshan distribution is phase-invariant, $\hat\rho=\int_\mathbb{C}d^2\alpha\,P(\alpha)|\alpha\rangle\langle\alpha|$ with $P(\alpha)=P(\alpha e^{i\varphi})$ for all $\varphi\in[0,2\pi[$.
	Namely, a phase-independent $P$ implies $\hat\rho=\int_0^\infty dr r P(r)\int_0^{2\pi}d\varphi\,|re^{i\varphi}\rangle\langle re^{i\varphi}|$, with $\alpha=re^{i\varphi}$ and $\int_0^{2\pi}d\varphi\,|re^{i\varphi}\rangle\langle re^{i\varphi}|=2\pi e^{-r^2}\sum_{n\in\mathbb N}r^{2n}|n\rangle\langle n|/n!$.
	Conversely, an incoherent mixture implies a phase-averaged $P$ distribution.
	To show this, we utilize that, given that $\hat\rho$ is incoherent, a uniform phase average (i.e., a fully dephasing channel) does not alter the state,
	\begin{equation}
	\begin{aligned}
		\hat\rho\
		=&\sum_{n\in\mathbb N}p_n|n\rangle\langle n|
		=\int_{0}^{2\pi}\frac{d\varphi}{2\pi}\,e^{i\varphi \hat n}\hat\rho e^{-i\varphi \hat n}
		\\
		=&\int_{0}^{2\pi}\frac{d\varphi}{2\pi}\int_{\mathbb C} d^2\alpha\,P(\alpha)|e^{i\varphi}\alpha\rangle\langle e^{i\varphi}\alpha|
		\\
		=&\int_{\mathbb C} d^2\alpha'\left[\int_{0}^{2\pi}\frac{d\varphi}{2\pi}\,P(e^{-i\varphi}\alpha')\right]|\alpha'\rangle\langle \alpha'|
	\end{aligned}
	\end{equation}
	using $e^{i\varphi \hat n}|\alpha\rangle=| \alpha'\rangle$, the substitution $\alpha'=e^{i\varphi}\alpha$, and the number operator $\hat n$, with $\hat n|n\rangle=n|n\rangle$.
	Therein, the Glauber-Sudarshan distribution of the state is $\int_{0}^{2\pi}d\varphi\,P(e^{-i\varphi}\alpha')/(2\pi)$, which is clearly phase-independent.

    More generally, one can show that all measures of quantum coherence and the circular variance under study are means to quantify the dephasing of a process. 
	Let $\Lambda$ be a generic dephasing channel, $\hat\rho^\mathrm{(out)}=\Lambda(\hat\rho^\mathrm{(in)})=\int d\varphi\,p(\varphi) e^{i\varphi\hat n}\hat\rho^\mathrm{(in)} e^{-i\varphi\hat n}$, where $p$ is a probability density.
	Applying the triangle inequality, we find 
	for the density matrix elements in the photon-number basis that
	\begin{equation}
	\begin{aligned}
		|\rho_{m,n}^\mathrm{(out)}|
		=&
		\left|
			\int d\varphi\,p(\varphi)e^{i\varphi(m-n)}\rho_{m,n}^\mathrm{(in)}
		\right|
		\\
		\leq&
		\int d\varphi\,p(\varphi)\left|
			\rho_{m,n}^\mathrm{(in)}
		\right|
		=|\rho_{m,n}^\mathrm{(in)}|
	\end{aligned}
	\end{equation}
	for $m,n\in\mathbb N$, showing a monotonic decrease of quantum coherence measures that are based on the off-diagonal elements in the photon-number expansion.
	In terms of the Glauber-Sudarshan distribution, we obtain $P^\mathrm{(out)}(\alpha)=\int d\varphi\,p(\varphi)P^\mathrm{(in)}(\alpha e^{-i\varphi})$ for the dephasing channel $\Lambda$, applying similarly to regularizations $P_\mathrm{\Omega}$.
	Now, we arrive at
	\begin{equation}
	\begin{aligned}
		&\left|
			\int d^2\alpha\,
			P^\mathrm{(out)}(\alpha)
			\frac{\alpha}{|\alpha|}
		\right|
		\\
		=&\left|
			\int d^2\alpha'\,
			P^\mathrm{(in)}(\alpha')
			\int d\varphi\,p(\varphi) e^{i\varphi}
			\frac{\alpha'}{|\alpha'|}
		\right|
		\\
		\leq&\left|
			\int d^2\alpha'\,
			P^\mathrm{(in)}(\alpha')
			\frac{\alpha'}{|\alpha'|}
		\right|
	\end{aligned}
	\end{equation}
	by substituting $\alpha'=\alpha e^{-i\varphi}$, also demonstrating a monotonically decreasing behavior, thus a monotonic increase of the circular variance.
	Thus, the dephasing induces a decrease of the amount of quantum coherence and an increase in the value of the circular variance, allowing one to quantify the dephasing with either quantity.

	The Glauber-Sudarshan distribution is typically not a well-behaved function, often including high singularities, rendering it an inconvenient quantity to assess the coherence via phase-space representations.
	To mitigate that deficiency, we can convolute $P$ with a sufficiently smooth kernel $\Omega$ so that we end up with a regularized $P_\Omega$ \cite{KV10},
	\begin{equation}
		\label{eq:convolution}
		P_{\Omega}(\alpha)=\int_{\mathbb C}d^2\gamma\,P(\gamma)\Omega(\alpha-\gamma).
	\end{equation}

	To avoid that the convolution introduces a phase bias (e.g., a phase dependence when $P$ is phase-independent), we can select a phase-invariant kernel $\Omega(\alpha)=\Omega(|\alpha|)$.
	Using Fourier methods, we show in the following that, in fact, such a kernel does not alter the phase properties.
	The Fourier transform of the Glauber-Sudarshan distribution, $P(\alpha)$, defines the characteristic function,
	\begin{equation}
		\label{eq:FT}
		\widetilde{P}(\beta)
		=\int_{\mathbb C} d^2\alpha\,
		P(\alpha) e^{\beta\alpha^\ast-\beta^\ast\alpha}.
	\end{equation}
	See Ref. \cite{VW06} for a thorough introduction of characteristic functions in quantum optics.
	The convolution in Eq. \eqref{eq:convolution} takes the form of a product in Fourier space, i.e.,
	\begin{equation}
		\widetilde P_{\Omega}(\beta)=\widetilde P(\beta)\widetilde \Omega(\beta).
	\end{equation}
	Importantly, the Fourier transform of the phase-invariant kernel is also phase-invariant,
	\begin{equation}
	\begin{aligned}
		\widetilde{\Omega}(\beta)
		=&\int_{\mathbb C} d^2\alpha\,
		\Omega(|\alpha|) e^{\beta\alpha^\ast-\beta^\ast\alpha}
		\\
		=&2\pi\int_{0}^\infty dr\,
		r\,
		J_0(2r|\beta|)
		\Omega(r)=\widetilde{\Omega}(|\beta|),
	\end{aligned}
	\end{equation}
	with $J_n$ denoting the $n$th Bessel function of the first kind.
	Hence, the phase information of $P$ is mapped onto the same information for the regularized $P_{\Omega}$, being essential to evaluate coherence properties.
	Please note that this result applies to non-Gaussian kernels $\Omega$, as we are using, as well as Gaussian ones, e.g., $\Omega(|\alpha|)= e^{-|\alpha|^2/\lambda}/(\pi\lambda)$ for $\lambda>0$, connecting the Glauber-Sudarshan distribution $P$ with the Wigner function $W$ and Husimi $Q$ function.

\subsection{Pattern functions for regularized phase-space distributions and a phase-invariant kernel}

	Now, we derive how we can reconstruct regularized phase-space distributions from data taken with balanced homodyne detectors by adapting the approach from Ref. \cite{KVHS11}.
	We begin with the Glauber-Sudarshan distribution, $P(\alpha)$.
	Its Fourier transform in Eq. \eqref{eq:FT} can be expressed as
	\begin{equation}
		\widetilde{P}(\beta)
		=\int_{\mathbb C} d^2\alpha\,
		P(\alpha) e^{\beta\alpha^\ast-\beta^\ast\alpha}
		=\langle e^{\beta\hat a^\dag}e^{-\beta^\ast \hat a}\rangle
	\end{equation}
	too \cite{VW06}, relating to the normally ordered displacement operator and where $\hat a$ and $\hat a^\dag$ are bosonic annihilation and creation operators.
	Applying the Baker-Campbell-Hausdorff formula, we can rewrite the operator in terms of quadrature operators $\hat q(\varphi)=e^{i\varphi}\hat a+e^{-i\varphi}\hat a^\dag$,
	\begin{equation}
		e^{\beta\hat a^\dag}e^{-\beta^\ast \hat a}
		=e^{|\beta|^2/2}e^{i|\beta|\hat q(\pi/2-\arg\beta)}.
	\end{equation}
	(Please note that the conjugate momentum to $\hat q(\varphi)$ is $\hat p(\varphi)=\hat q(\varphi+\pi/2)$ such that $[\hat q(\varphi),\hat p(\varphi)]=2i$ holds true.)
	Via the inverse Fourier transform, we find for the $P$ distribution the expression
	$
		P(\alpha)=\pi^{-2}\int_{\mathbb C}d^2\beta
		e^{\beta^\ast\alpha-\beta\alpha^\ast}
		e^{|\beta|^2/2}
		\langle e^{i|\beta|\hat q(\pi/2-\arg\beta)}\rangle
	$.
	The substitution $\beta=ise^{-i\varphi}$ with $s\in\mathbb R$ and $0\leq \varphi<\pi$ yields
	\begin{equation}
		P(\alpha)=
		\int_{-\infty}^\infty ds\,
		\int_0^\pi d\varphi\,
		\frac{|s|}{\pi^2}
		e^{-2is\mathrm{Re}(e^{i\varphi}\alpha)}
		e^{s^2/2}
		\langle e^{is\hat x(\varphi)}\rangle.
	\end{equation}

	The quadrature operator can be given in its spectral decomposition, i.e., $\hat x(\varphi)|x;\varphi\rangle=x|x;\varphi\rangle$, resulting in
	\begin{equation}
	\begin{aligned}
		P(\alpha)=&
		\int_{-\infty}^\infty ds\,
		\int_0^\pi d\varphi\,
		\frac{|s|}{\pi^2}
		e^{-2is\mathrm{Re}(e^{i\varphi}\alpha)}
		e^{s^2/2}
		\\
		&\times\int_{-\infty}^\infty dx\,
		e^{isx}
		\langle x;\varphi |\hat\rho|x;\varphi \rangle
		\\
		=&\int_{-\infty}^\infty dx\,
		\int_0^\pi d\varphi\,
		p(x;\varphi)f(\alpha;x;\varphi),
	\end{aligned}
	\end{equation}
	for a density operator $\hat\rho$, measured quadrature probabilities $p(x;\varphi)=\langle x;\varphi|\hat\rho|x;\varphi\rangle$, and
	\begin{equation}
		f(\alpha;x;\varphi)=\int_{-\infty}^\infty ds \frac{|s|}{\pi^2}
		e^{-2is\mathrm{Re}(e^{i\varphi}\alpha)}
		e^{s^2/2}
		e^{isx},
	\end{equation}
	defining the divergent pattern function for the Glauber-Sudarshan distribution.
	Note that $p(x;\varphi)$ is the probability density for measuring $x$ for a fixed angle $\varphi$, i.e., the conditional probability $p(x;\varphi)=p(x,\varphi)/p(\varphi)$, with the joint and marginal (w.r.t. the angle) densities $p(x,\varphi)$ and $p(\varphi)$.

	In order to regularize the above expression, we can consider a convoluted Glauber-Sudarshan distribution, Eq. \eqref{eq:convolution}.
	This yields the pattern function
	\begin{equation}
	\begin{aligned}
		f_\Omega(\alpha;x;\varphi)
		=&
		\int_{\mathbb C} d^2\gamma\,\Omega(\gamma)
		\int_{-\infty}^\infty ds\,
		\frac{|s|}{\pi^2}
		\\
		&\times
		e^{-2is\mathrm{Re}(e^{i\varphi}[\alpha-\gamma])}
		e^{s^2/2}
		e^{isx}.
	\end{aligned}
	\end{equation}
	We can now utilize a suitable regularizing function  \cite{KV12a,KV12b},
	\begin{equation}
		\label{eq:ChosenKernel}
		\Omega(\gamma)=\left[\frac{J_1(2R|\gamma|)}{\sqrt{\pi}|\gamma|}\right]^2.
	\end{equation}
	In polar coordinates, $\gamma=re^{i\vartheta}$, this yields
	\begin{equation}
	\begin{aligned}
		f_\Omega(\alpha;x;\varphi)
		=&
		\int_{0}^{\infty} dr\,
		\int_{0}^{2\pi} d\vartheta\,
		\int_{\mathbb R} ds\,
		\frac{|s|}{\pi^3 r}
		J_1(2Rr)^2
		\\
		&\times
		e^{2isr\cos(\varphi+\vartheta)}
		e^{is[x-2\mathrm{Re}(e^{i\varphi}\alpha)]}
		e^{s^2/2}
		\\
		=&
		\int_{\mathbb R} ds\,
		\frac{2|s|}{\pi^2}
		\int_{0}^{\infty} dq\,
		\frac{J_1(q)^2
		J_0\left(\frac{|s|}{R}q\right)}{q}
		\\
		&\times
		e^{is[x-2\mathrm{Re}(e^{i\varphi}\alpha)]}
		e^{s^2/2},
	\end{aligned}
	\end{equation}
	applying a transformation $q=2Rr$.
	It is now convenient to evaluate the transformed integral,
	\begin{equation}
	\begin{aligned}
		g(t)=&
		\int_{0}^{\infty} dq\,
		\frac{J_1(q)^2
		J_0\left(2tq\right)}{q}
		\\
		=&\frac{1}{\pi}
		\left\lbrace\begin{array}{lll}
				0 & \text{ for } & t>1,
				\\
				\arccos(t)-t \sqrt{1-t^2} & \text{ for } & 0\leq t\leq 1,
		\end{array}\right.
	\end{aligned}
	\end{equation}
	where $2t=|s|/R$.
	This allows us to express the pattern function via an integral over a finite interval,
	\begin{equation}
	\begin{aligned}
		f_\Omega(\alpha;x;\varphi)
		=&
		\int_{-2R}^{2R} ds\,
		\frac{2|s|}{\pi^2}
		g\left(\frac{|s|}{2R}\right)
		e^{is[x-2\mathrm{Re}(e^{i\varphi}\alpha)]}
		e^{s^2/2}
		\\
		=&\frac{16R^2}{\pi^3}h(X,R),
	\end{aligned}
	\end{equation}
	where $X=2R[x-2|\alpha|\cos(\varphi+\arg\alpha)]$ and, for $u=|s|/(2R)$,
	\begin{equation}
	\begin{aligned}
		h(X,R)=\int_0^{1} du\,&
		u\left[
			\arccos(u)-u\sqrt{1-u^2}
		\right]
		\\
		&\times
		\cos(uX)e^{2R^2u^2},
	\end{aligned}
	\end{equation}
	which can be evaluated numerically for a given $R$ and any $X$.

	All above considerations enable us to analytically write
	\begin{equation}
	\begin{aligned}
		&P_{\Omega}(\alpha)
		=
		\int_{-\infty}^\infty dx\,
		\int_0^\pi d\varphi\,
		p(x;\varphi)f_{\Omega}(\alpha;x;\varphi)
		\\
		=&
		\int_{-\infty}^\infty dx\,
		\int_0^\pi \frac{d\varphi}{\pi}\,
		\frac{p(x,\varphi)}{\int_{\mathbb R} dx'\,p(x',\varphi)}\pi f_{\Omega}(\alpha;x;\varphi).
	\end{aligned}
	\end{equation}
	Now suppose $N$ quadratures $x_i$ are measured for the phase $\varphi_i$, i.e., the data set reads $\{(x_i,\varphi_i)\}_{i=1,\ldots,N}$.
	Then, the above formula can be replaced as described in the following.
	Say $N(x,\varphi)$ is the number of data points $(x_i,\varphi_i)$ which are closest---according to the defined grid of $\Xi$ position intervals and $\Phi$ phase intervals---to $(x,\varphi)$.
	Then the above integral is approximated through the weighted average
	\begin{equation}
		P_{\Omega}(\alpha)\approx
		\frac{1}{\Phi}\sum_{x,\varphi} \frac{N(x,\varphi)}{\sum_{x'}N(x',\varphi)}\pi f_{\Omega}(\alpha;x;\varphi)=\overline{P_{\Omega}(\alpha)}.
	\end{equation}
	The second-order moment $\overline{P_{\Omega}(\alpha)^2}$ can be defined similarly, using $[\pi f_{\Omega}(\alpha;x;\varphi)]^2$ in the above expression.
	Finally, the uncertainty for each phase-space point is directly given as
	\begin{equation}
		\sigma[P_{\Omega}(\alpha)]=\sqrt{\frac{\overline{P_{\Omega}(\alpha)^2}-\overline{P_{\Omega}(\alpha)}^2}{N-1}}.
	\end{equation}

	As a final comment, it is worth mentioning that altering the parameter $R>0$ that defines the sampling function $f_{\Omega}$ allows one to get various phase-space distributions.
	For $R\to\infty$ (wide filter), we approach the original Glauber-Sudarshan distribution \cite{KV10}.
	However, this also leads to large errors $\sigma$.
	Conversely, $R\to0$ (narrow filter), does not provide enough information to significantly estimate the phase-space distribution.
	Thus, the Goldilocks zone is to be found for intermediate $R$ values;
	see, e.g., Ref. \cite{ASVKMH15} and Sec. \ref{supp.sec:Implementation}.


\section{Simulation of regularized $P$ function dynamics}\label{supp.sec:numerics}

    Here, we describe our numerical model, as used in Ref. \cite{LPRSHSSA21}.
    Importantly, we extend the existing approach that is based on Wigner functions $W$ to a new one that is compatible with our regularized $P$ functions.

\subsection{Wigner function evolution}

    We use an open-dissipative Gross-Pitaevskii equation coupled to an incoherent reservoir in which fluctuations are taken into account via a truncated Wigner approximation (TWA).
    Then sought-after properties can be calculated using Monte Carlo techniques.
    See Refs. \cite{SLC02,BBD08,WS09,CC12} for introductions.
    The main idea is to employ the Wigner function for a bosonic polariton field operator $\hat{\psi}(\mathbf{r})$ that can be used to sample the phase-space distribution $W(\psi (\mathbf{r}))$, where $\alpha=\psi (\mathbf{r})$ in comparison to Sec. \ref{supp.sec:regP}.
    The TWA then describes time evolution of the Wigner distribution as a set of stochastic partial differential equations for the corresponding phase-space variables.

    On a finite two-dimensional spatial $N\times N$ grid of length $L$ ($\mathbf{r}\equiv\mathbf{r}_i$, $i\in\lbrace 1,...,N^2\rbrace$ denoting the grid points), the dynamics of the polariton field $\psi(\mathbf{r})$ when coupled to the incoherent reservoir density $n_\mathrm{res}(\mathbf{r})$ reads \cite{WS09}
    \begin{equation}\label{eq:num_model}
	\begin{aligned}
        d\psi=& \frac{1}{i\hbar}\bigg[H+\frac{\mathrm{i}\hbar}{2}\left( R_r n_\mathrm{res}-\gamma_c \right)+g_r n_\mathrm{res}+g_c|\psi|_-^2 \bigg]
        \\
        &\times \psi\, dt+dW_\mathrm{noise},
        \\
        \frac{\partial n_\mathrm{res}}{\partial t}=&\left(-\gamma_r-R_r |\psi|_-^2 \right)n_\mathrm{res}+P_\mathrm{pump}.
	\end{aligned}
    \end{equation}
    Therein, $H=-\hbar^2\nabla^2/(2m_{\mathrm{eff}})$ is the free-particle Hamiltonian, the effective polariton mass is $m_{\mathrm{eff}}$,
    $\gamma_c$ and $\gamma_r$ are the decay rates of the condensate and the reservoir, the interaction strength between polaritons is $g_c$, the condensation rate and condensate-reservoir interaction are $R_r$ and $g_r$, and $P_\mathrm{pump}(\mathbf{r})=P_0 \exp\left[-\mathbf{r}^2/w^2\right]$ with $w=40~\mu\mathrm{m}$ (matching the FWHM of $70~\mu\mathrm{m}$ used in the experiment) is the continuous-wave Gaussian pump profile.
    Similar to the experiment, we define the threshold excitation power $P_{\mathrm{thr}}$ via the onset of quantum coherence in the selected mode;
    explicitly $P_{\mathrm{thr}}=8~\mathrm{ps}^{-1}\mu\mathrm{m}^{-2}$ for the parameter choice defined below.
    In addition, the renormalized condensate density reads $|\psi|^2_-{\equiv}|\psi|^2{-}(\Delta V)^{-1}$, with the unit cell volume $\Delta V=L^2/N^2$.
    The complex-valued Wiener noise contribution is given by $dW_\mathrm{noise}$, with correlations satisfying
    \begin{equation}
    \begin{aligned}
        \langle dW_\mathrm{noise}(\mathbf{r})dW_\mathrm{noise}(\mathbf{r'}) \rangle &=0
        \quad\text{and}
        \\
        \langle dW_\mathrm{noise}(\mathbf{r})dW_\mathrm{noise}^\ast(\mathbf{r'}) \rangle
        &=\left( R_r n_\mathrm{res}+\gamma_c \right)\frac{\delta_{\mathbf{r},\mathbf{r'}}\,dt}{2\Delta V}.
    \end{aligned}
    \end{equation}
    
    The following system parameters are used in the numerical simulations $m_{\mathrm{eff}}=10^{-4} m_e$ ($m_e$ is the electron mass), $\gamma_c=0.2~\mathrm{ps}^{-1}$, $\gamma_r=1.5 \gamma_c$, $R_r=0.015~\mathrm{ps}^{-1}~\mu\mathrm{m}^2$, $g_c=g_r=6\times 10^{-3}~\mathrm{meV}~\mathrm{\mu m}^2$.
    A very similar choice of parameters has been established in our previous work to describe the same semiconductor microcavity sample \cite{MBADMSHS20,LPRSHSSA21}.
    However, here we use a much smaller value for the condensate-reservoir interaction $g_r$.
    While the amount of quantum coherence is not very sensitive to changes of $g_r$, here we found that it does heavily influences the coherence time through the nonlinear part of the polariton's effective potential landscape, as discussed in the main part.
    The numerical model is solved with a fourth-order stochastic Runge--Kutta algorithm \cite{HP06} on a finite two-dimensional grid in real space with box width $L=230.4\,\mu\mathrm{m}$ and a step size of $\sqrt{\Delta V}=0.9\,\mu\mathrm{m}$.
    Further details on the numerical implementation can be found in Ref. \cite{LPRSHSSA21}.
    
\subsection{Initial state preparation and final mode selection}

    An important aspect of studying the decay of phase coherence in the time domain is how the phase-space data is obtained both in the experimental realization and the numerical simulation. 
    In the latter, we use the following recipe to prepare the initial states which are assumed as displaced thermal states, i.e. they can be described by a Gaussian phase-space distribution:
    (i) For a selected single mode $\mathbf{k}_s$, the steady-state expectation values for mean $\langle \hat{n}_{\mathbf{k}_s} \rangle$, variance $\langle (\Delta\hat{n}_{\mathbf{k}_s})^2 \rangle$, and the quantum coherence $\mathcal{C}$ are calculated as described in detail in Ref. \cite{LPRSHSSA21}. 
    These three values are the target values for the initial state preparation.
    (ii) We determine the solution's spatial envelop via a mean-field model;
    see Eq. \eqref{eq:num_model} for $dW_\mathrm{noise}=0$ and $|\psi|_-^2=|\psi|^2$.
    (iii) For each spatial grid point, we draw phase-space samples from a normal distribution with mean $\mu$ and standard deviation $\sigma$. 
    These sampled grids are then multiplied by the normalized spatial envelope from the previous step.
    (iv) For this artificially generated state, expectation values for mean, variance and quantum coherence of the selected single mode $\mathbf{k}_s$ are calculated and compared to the target values.
    (v) Depending on the deviation, $\mu$ and $\sigma$ are varied and the procedure is repeated from step (iii) until a desired agreement is obtained.
    Now, this state yields the correct steady-state expectation values for mean, variance, and quantum coherence and, at the same time, has the right amount of phase coherence corresponding to a Gaussian phase-space distribution.
    Now, the decay of coherence can be dynamically tracked via the circular phase variance of the phase-space distribution.
    For the numerical results presented in the main part, we choose the single mode $\mathbf{k}_s=\mathbf{0}$.
    However, choosing neighboring modes within the condensate's finite $k$-space emission spot yields very similar results.
    In principle, the method can be extended to multi-mode selection at the expanse of increased computational effort.

\begin{widetext}

\subsection{Relation to regularized phase-space distributions}

	By applying the aforementioned approach, we can simulate the evolution of the Wigner function.
	A new method that we establish here is utilizing this dynamics to assess the evolution of regularized $P_\Omega$ functions.

	To this end, we again consider characteristic functions.
	The transition from $\widetilde W(\beta)$ for the Wigner function to $\widetilde P(\beta)$ for the Glauber-Sudarshan distribution is given by $\widetilde P(\beta)=e^{|\beta|^2/2}\widetilde W(\beta)$.
	The transformation from $\widetilde P(\beta)$ to the characteristic function $\widetilde P_\Omega$ of the regularized phase-space function can be described via the autocorrelation function of a (normalized) cylinder, i.e., the Fourier transform of the kernel in Eq. \eqref{eq:ChosenKernel}.
	Combining these relations and computing the inverse Fourier transform yields the convolution kernel $K(\alpha)$ that maps the Wigner function $W(\alpha)$ to $P_\Omega$ for all times.
	We have
	\begin{equation}
	\begin{aligned}
		K(\alpha)
		=&\int \frac{d^2\beta}{\pi^2}e^{\beta^\ast\alpha-\beta\alpha^\ast}e^{|\beta|^2/2}
		\int \frac{d^2\gamma}{\pi R^2}\Theta(R-|\gamma|)\theta(R-|\gamma+\beta|)
		\\
		=&\frac{1}{\pi^3 R^2}\int_{|\gamma|\leq R} d^2\gamma\int_{|\delta|\leq R}d^2\delta\,
		\exp\left[
			(\delta-\gamma)^\ast\alpha
			-(\delta-\gamma)\alpha^\ast
			+\frac{|\delta-\gamma|^2}{2}
		\right]
		\\
		=&\frac{1}{\pi^3 R^2}\int_0^R ds\,s\int_0^R dt\,t \int_0^{2\pi}d\varphi\int_0^{2\pi}d\vartheta\,
		\exp\left[
			-2i|\alpha|t\sin(\vartheta)
			+2i|\alpha|s\sin(\varphi)
			+\frac{s^2+t^2-2st\cos(\varphi-\vartheta)}{2}
		\right]
		\\
		=&\frac{R^2}{\pi^3}\int_0^1 ds'\,s'\int_0^1 dt'\,t' \int_0^{2\pi}d\varphi\int_0^{2\pi}d\vartheta\,
		\exp\left[
			-iA\Big(t'\sin(\vartheta)
			-s'\sin(\varphi)\Big)
			+\frac{R^2}{2}\Big(t^{\prime 2}+s^{\prime 2}-2s't'\cos(\varphi-\vartheta)\Big)
		\right],
	\end{aligned}
	\end{equation}
	using the substitutions $\delta=\gamma+\beta$, $\gamma=s e^{i(\varphi+\arg\alpha)}$, $\delta=t e^{i(\vartheta+\arg\alpha)}$, $s'=s/R$, $t'=t/R$, and $A=2R|\alpha|$.

	The resulting integral over finite domains can be evaluated numerically to obtain the sought-after kernel.
	We then carry out a convolution of the Wigner function at different times with this kernel to obtain our regularized $P_\Omega$ for that times.
	Finally, the circular phase variance of the reconstructed $P_\Omega$ is calculated and can be directly compared to the experimental results.

\end{widetext}


\begin{figure*}
	\includegraphics[width=0.7\textwidth]{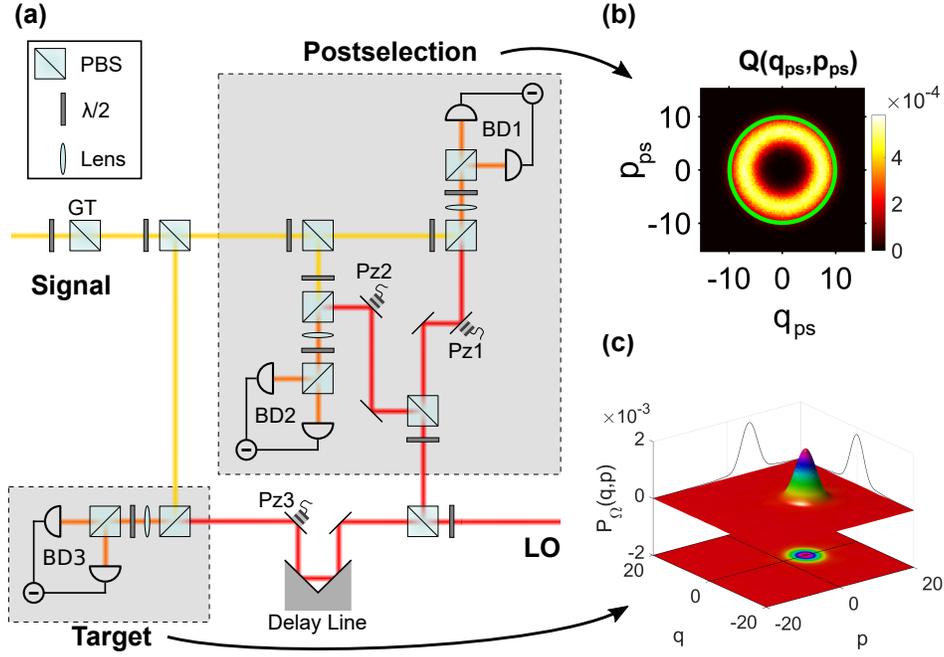}
	\caption{%
		Experimental setup (a).
		The signal from the polariton microcavity is split into three channels.
		Each channel is interfered with the local oscillator (LO), before being detected by a balanced homodyne detector (BD).
		While BD1 and BD2 are placed in the postselection arm in order to measure the Husimi $Q$ function (b) of the signal light field, BD3 is placed in the target arm which is used to reconstruct the regularized $P$ function (c).
		The green ring in (b) exemplifies one selected phase-space region.
		(PBS: polarizing beam splitter; GT: Glan-Thompson prism; $\lambda/2$: half-wave plate; Pz: piezo mirror.)
	}\label{supp.fig:Setup}
\end{figure*}

\begin{figure*}
    \includegraphics[width=0.7\textwidth]{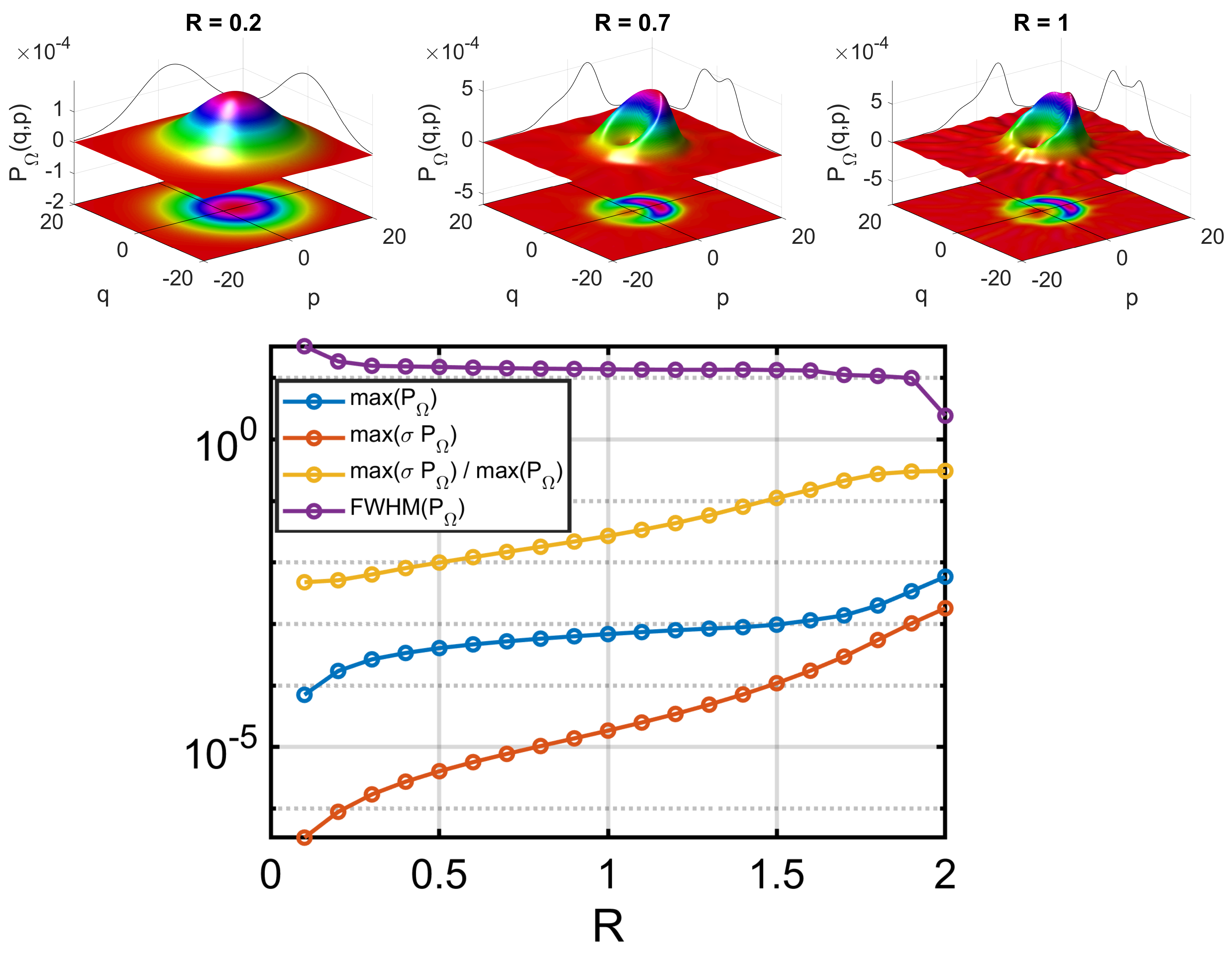}
    \caption{%
		Comparison of the effect of different filter parameters $R$.
		The depicted data belong to $P_{\text{exc}} = 400~\mathrm{mW}$ ($P_{\text{exc}} = 1.7~P_\mathrm{thr}$) and a time delay $\tau = 1214~\mathrm{ps}$.
		With increasing $R$, the $P$ function evolves from a broad, featureless shape with a small error, to a more wrinkled shape with higher error.
		The value $R = 0.7$ yields a good trade-off between well-defined features and reasonable reconstruction errors.
    }\label{supp.fig:Rcomparison}
\end{figure*}

\begin{figure*}
    \includegraphics[width=1\textwidth]{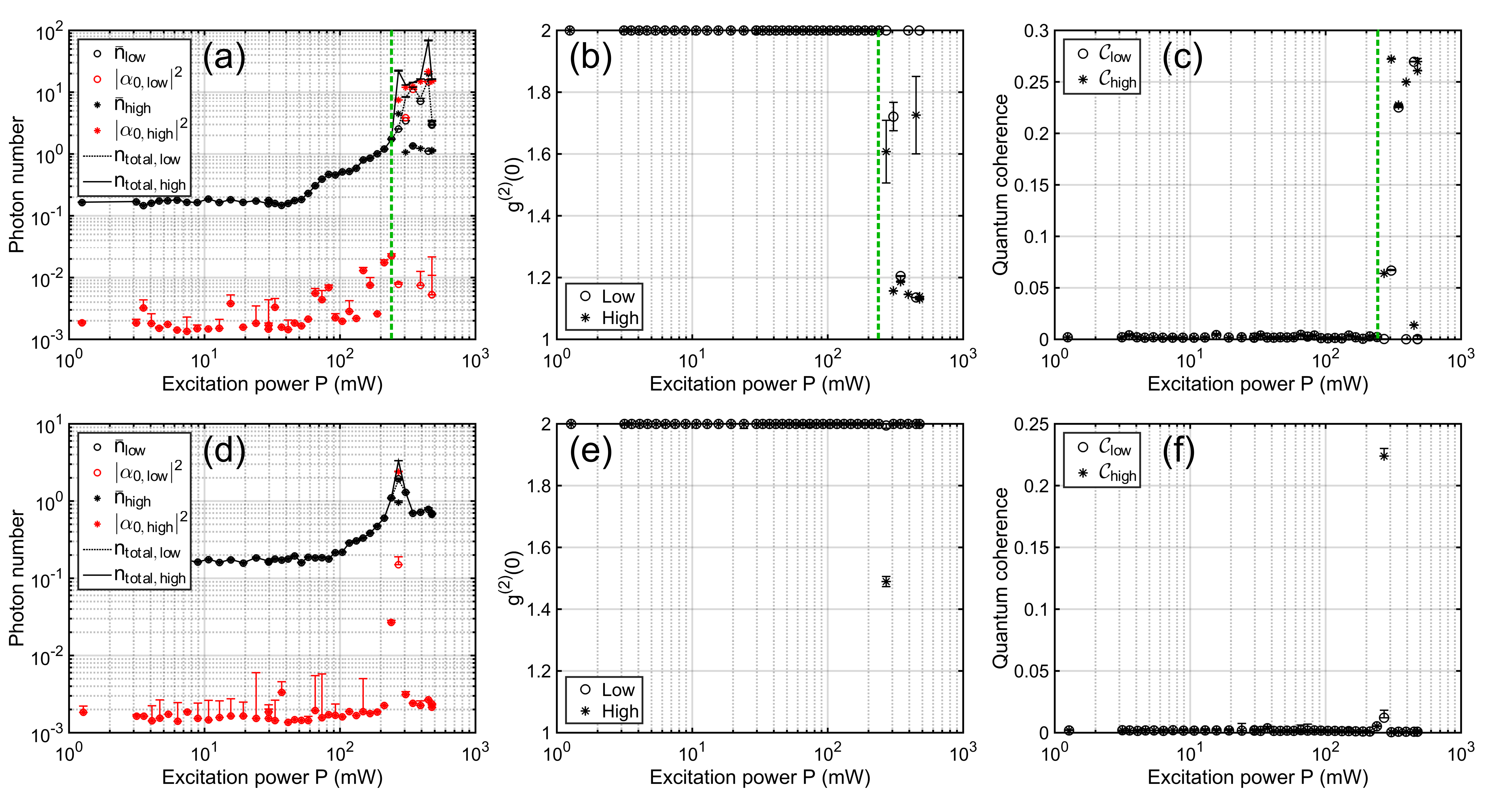}
    \caption{%
		Power dependence for the dominant linear polarization (upper row) and its orthogonal complement (lower row), extracted from homodyne detection measurements.
		The green dashed lines in (a)--(c) indicate the thresholds for a nonlinear increase of the coherent photon number and the onset of quantum coherence in the dominant polarization. 
		Plots (a) and (d): Coherent (red circles) and thermal (black circles) photon numbers.
		Black line shows the total photon number $n_{\mathrm{total}}$. 
		Plots (b) and (e): Equal-time, second-order correlation function $g^{(2)}(\tau=0)$.
		Plots (c) and (f): Amount of quantum coherence.
		Open and closed symbols correspond to the low and high state, respectively, when the emission is switching on and off.
		Both symbols overlap when the emission is stable.
		Because of the asymmetry of the logarithmic scale, only the upper part of the error margin is depicted.
    }\label{supp.fig:excPower}
\end{figure*}

 \begin{figure*}
    \includegraphics[width=0.8\textwidth]{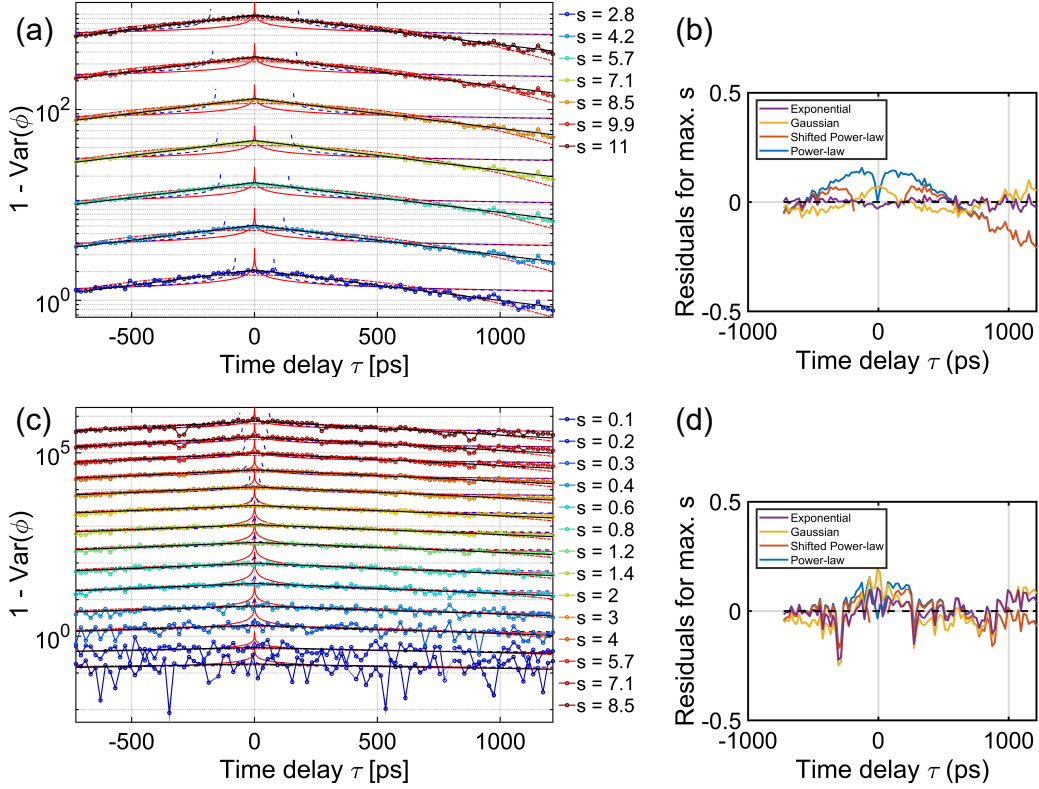}
    \caption{%
		Comparison of different fit functions.
		Plots (a) and (b): $P_{\text{exc}} = 1.7~P_\mathrm{thr}$.
		The width $w$ of the selected phase-space region is $0.57$.
		(a) Temporal decays of $1-\mathrm{Var}(\phi)$, plotted semilogarithmically and scaled.
		(Solid red line: power-law fit function;
		dashed blue line: temporally shifted power law; dotted red line: Gaussian fit; solid black line: exponential fit.)
		(b) Residuals (i.e., data minus fit) for the highest postselection radius.
		Plots (c) and (d): $P_{\text{exc}} = 1~P_\mathrm{thr}$.
        The width $w$ is $0.1$, except for the three highest radii $s$, where $w = 0.57$.
    }\label{supp.fig:fitfunctions}
\end{figure*}


\section{Additional details on the experimental implementation}\label{supp.sec:Implementation}

	In this section, details about our experiment are provided.
	This includes technical properties of our setup and sample and lays out our data processing and analysis methods.

\subsection{Sample, signal, and setup}

    Our polariton microcavity consists of two distributed-Bragg reflectors with alternating layers of Al$_{0.2}$Ga$_{0.8}$As and AlAs, featuring a high quality factor of about $20\,000$. 
    Between these mirrors lies a $\lambda/2$ cavity, containing four GaAs quantum wells.
    The Rabi splitting of the sample is $9.5~\mathrm{meV}$.
    In our experiments, we chose an exciton cavity detuning of $-6.2~\mathrm{meV}$.
    A detailed sample characterization can be found in Refs. \cite{MBADMSHS20,LPRSHSSA21}.

    The sample is pumped nonresonantly with a continuous-wave laser at the first minimum of the stop band, having a Gaussian spatial profile with a large diameter of $70~\mathrm{\mu m}$ full width at half maximum (FWHM), in order to favor condensation at $k = 0$ \cite{KRKBJKMSASSLDD06,WCC08}.
    The emission is spatially filtered by a pinhole with $100~\mathrm{\mu m}$ diameter, corresponding to $13~\mathrm{\mu m}$ on the sample, to ensure the homogeneity of the emission in the investigated region.
    This also corresponds to a filtering in $k$ space around $k = 0$ with a FWHM of $1~\mathrm{\mu m}^{-1}$.

    The polarization of the emission can be chosen with a half-wave plate and a Glan--Thompson prism.
    For this work, only the most dominant linear polarization was measured.
    
	A setup overview is shown in Fig. \ref{supp.fig:Setup}.
    The signal from the polariton microcavity is split into three channels.
    Each channel is interfered with the local oscillator (LO).
    The LO, derived from a pulsed Ti:sapphire laser, is resonant with the most intense zero-momentum ground-state mode of the polariton emission, with a FWHM of $1.9~\mathrm{nm}$.
    The LO has a Gaussian spatial mode and a FWHM in $k$ space of $1.3~\mathrm{\mu m}^{-1}$, centered at $k = 0$.
    Therefore, the LO overlaps only with signal components around $k \approx 0$.

    In the three channels, the field quadratures of the emission $X_1$, $X_2$, and $X_3$ are measured via homodyne detection through the balanced detectors BD1, BD2 and BD3 (Fig. \ref{supp.fig:Setup}).
    Since the LO is pulsed with a repetition rate of $75.4~\mathrm{MHz}$, each pulse delivers one quadrature measurement.
    During measurement, the relative phase between the channels is swept continuously by a piezo mirror in the LO path in channel 1.
    A time delay $\tau$ between the target channel and the other channels can be set with a delay line. 
    Further technical details regarding our detection system can be found in Ref. \cite{TLA20}.	
    
\subsection{Data processing}    

    In postprocessing, we remove correlations between consecutive quadrature measurements \cite{KBMCHL12} and normalize the result with respect to the LO amplitude in order to obey the commutator convention $[\hat q,\hat p]=2i$ for a quadrature $q$ and its conjugate momentum $p$ (cf. Sec. \ref{supp.sec:regP}) such that the fluctuation for vacuum yields $\Delta q_{\text{vac}} = 1$. 
    We can also remove a slow drift of the mean photon number over the course of the delay series and only utilize data where the photon number is stable inside a given range during one measurement, removing jumps outside that range.
    We ensure that the quadratures of channel 1 and 2 are orthogonal---to provide a measurement of the Husimi function---by keeping only data where the product $X_{1} \times X_{2}$ equals zero within a $\pm2.5\%$ margin of the peak-to-peak value.
    Then, the quadratures $X_1$ and $X_2$ represent the phase-space coordinates $q_{ps}$ and $p_{ps}$.
    And we assemble a histogram of the occurrences of pairs $(q_{ps},p_{ps})$, resulting in the Husimi $Q$ distribution \cite{S01}.

    Additionally, we reconstruct the relative phase between the signal in the target channel and the LO.
    This consists of two components: the relative phase in the postselection arm $\varphi_{ps} = \arctan(p_{ps}/q_{ps})$, with the four-quadrant inverse tangent function, and the relative phase $\Delta \varphi$ between the LOs in postselection and target channel.
    The latter phase is derived from the smoothed time dependence of $X_1 \times X_3$, which takes a sinusoidal form due to the piezo mirror modulation.
    From these partial phases, the relative phase between signal and LO in the target channel $\varphi$ is computed via $\varphi =(\Delta \varphi + \varphi_{ps})\,\mathrm{mod}\, 2\pi$.

    Altogether, we obtain a data set $(q_{ps},p_{ps},X_3,\varphi)$ for each LO pulse.
    Now, we select a specific phase-space region of the Husimi function, specified by a radius $s$ and a thickness $w$.
    An example region is depicted in green in Fig. \ref{supp.fig:Setup}(b).
    Then, we pick those target quadrature values $q \in X_3$, which have been measured at the same instant as the data in the selected region, and the corresponding phases $\varphi$.

    With this dataset $(q,\varphi)$, we then reconstruct the regularized $P$ function via the pattern function scheme described in Sec. \ref{supp.sec:regP}.
    We chose a grid of quadrature and phase intervals as $x_{\text{grid}} \in [-20,20]$ with equally spaced step size $1$, and $\varphi_{\text{grid}} \in [0,2\pi]$ with equally spaced step size $0.1$.
    For the resulting phase-space coordinates over which the $P$ function is displayed, we chose $q,p \in [-20,20]$ with step size $0.25$, where $\alpha=q+ip$ defines the complex phase-space variable.
    All these step sizes were chosen as a trade-off between the smoothness of the regularized $P$ function and a reasonable computation time.

    A filter parameter $R = 0.7$ has been chosen for the regularized $P$ function.
    A comparison of different $R$ values is shown in Fig. \ref{supp.fig:Rcomparison}.
    Too small filter parameters lead to too broad $P$ functions, washing out important features, while too high $R$ values cause increased uncertainties.
    We found that $R = 0.7$ yields a good resolution, while avoiding too large error margins.
    
\subsection{Initial sample characterization}

	From measuring the Husimi function, we can derive the coherent and thermal photon numbers, the second-order photon correlation $g^{(2)}(\tau = 0)$, and the quantum coherence;
	see Ref. \cite{LPRSHSSA21} for a detailed explanation.
	Figure \ref{supp.fig:excPower} shows these quantities as a function of the excitation power.
	For the dominant polarization, we observe a rapid increase of coherent photon number and quantum coherence and a corresponding decrease of $g^{(2)}(\tau = 0)$ at $P_\mathrm{exc} > 240~\mathrm{mW}$.
	Therefore, we define this value as the threshold power $P_\mathrm{thr}$ for our purpose.
	(A spectroscopic measurement reveals a nonlinear increase of intensity at even lower powers, but this does not coincide with an increase of coherence in the measured mode due to mode competition.)
	Between $P_\mathrm{exc} = 240~\mathrm{mW}$ and $P_\mathrm{exc} = 270~\mathrm{mW}$, there is also a significant coherent photon number for the orthogonal polarization, which vanishes for higher powers.
	It is reasonable to assume that, in this power range, mode competition occurs between the two polarizations.

\subsection{Fit functions for temporal decays}

	Different fit functions are compared to describe the temporal decays of the circular variance.
	Figure \ref{supp.fig:fitfunctions}(a) shows for $P_{\text{exc}} = 1.7~P_\mathrm{thr}$ the temporal decay of $1-\mathrm{Var}(\phi)$ on a semilogarithmic scale so that different functions are easily distinguishable by their shape.
	The data for increasing postselection radii were multiplied by an exponentially increasing factor in order to make all of them equally visible in this depiction.
	To these data, we fitted a power-law decay, a power-law decay that starts at variable time, a Gaussian decay, and an exponential decay.
	Figure \ref{supp.fig:fitfunctions}(b) depicts the residual difference between data and fit for the highest radius.
	The exponential fit function presents the best match for our data.

	A similar result was found for all other excitation powers, except $P_{\text{exc}} = 1~P_\mathrm{thr}$.
	At this specific power, shown in Fig. \ref{supp.fig:fitfunctions}(c) and (d), the decay of $1-\mathrm{Var}(\phi)$ exhibits a flatter slope, especially for larger time delays, and a power-law decay also seems applicable although the improvement compared to the exponential fit is not decisive.

	In theory, a power-law decay was suggested for large systems and intermediate times \cite{SKL07} and experimentally observed as an indicator of a Berezinskii--Kosterlitz--Thouless phase transition for a high-quality sample with a spatially extended condensate \cite{CBDSGDWPGLSS18}.
	This kind of topological order might persist at the threshold power because here the vortex-antivortex pairs are less disrupted by interactions than at higher powers.
	Overall, however, we chose an exponential fit to determine the decay times for all excitation powers and for the temporal decays of the mean amplitude. \\

\begin{figure}
    \includegraphics[width=0.48\textwidth]{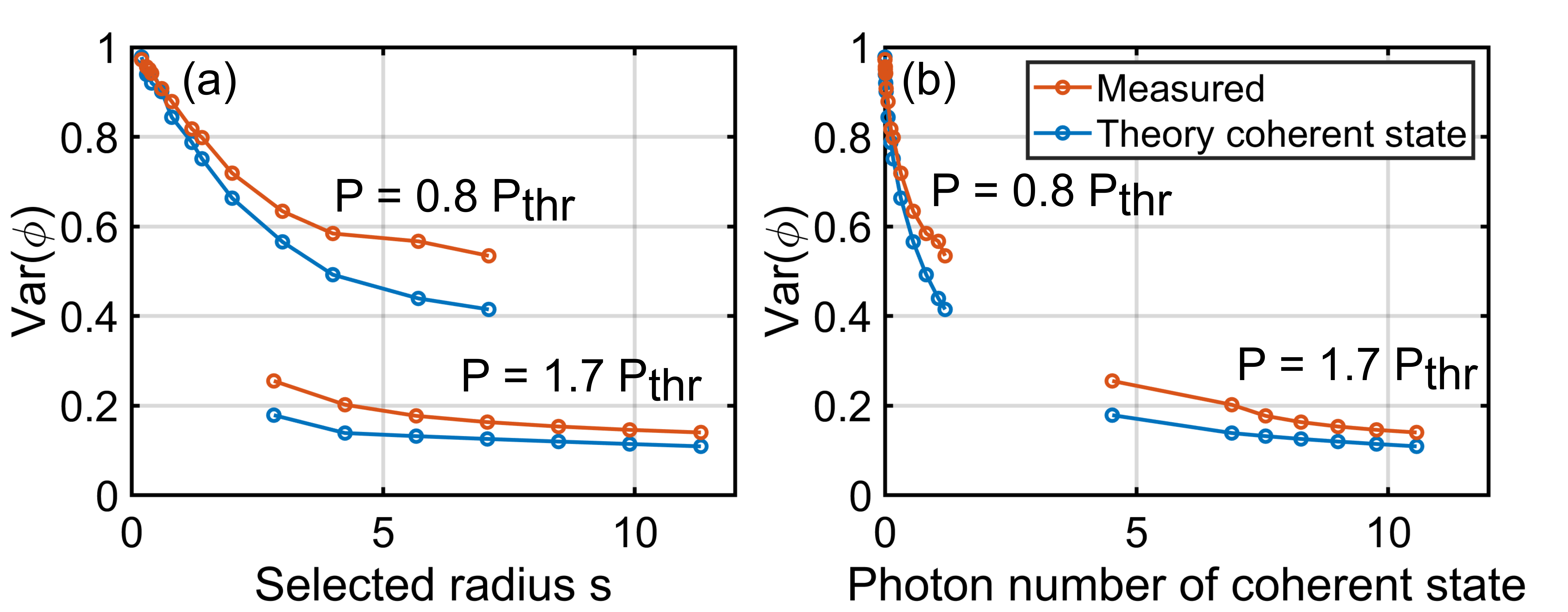}
    \caption{%
		(a) Circular phase variance $\mathrm{Var}(\phi)$ versus the selected radius $s$ for $P_{\text{exc}} = 0.8~P_\mathrm{thr}$ and $P_{\text{exc}} = 1.7~P_\mathrm{thr}$.
		Red: Measured $\mathrm{Var}(\phi)$ around zero time delay.
		Blue: $\mathrm{Var}(\phi)$ of a theoretical coherent state with an amplitude corresponding to the postselected state.
		(b) Measured (red) and theoretical (blue) circular phase variance $\mathrm{Var}(\phi)$ versus the photon number of the corresponding coherent state.
    }\label{supp.fig:MinPhaseVar}
\end{figure} 

\subsection{Remarks on the minimum phase variance}

    In Fig. 2 of our paper, we observe an increase of the circular phase variance around zero delay for decreasing selected radii $s$.
    This behavior can be attributed to the fundamental photon number-phase uncertainty relation \cite{SBCR93}
    \begin{equation}
        \Delta \phi \Delta n \geq \frac{1}{2}|\langle [\hat{\phi},\hat{n}] \rangle|.
    \end{equation}
    When choosing a smaller radius $s$, one has a state with a lower photon number, leading to a higher phase uncertainty.
    This can be verified when comparing the measured phase variance to the one of a theoretical coherent state with the same amplitude as the postselected state.
    To this aim, we calculate the $P_{\Omega}$ function of a coherent state that has the same offset $q_{0}$ as the $P_{\Omega}$, which was reconstructed from the measured data, and evaluate the coherent state's circular phase variance;
    see Fig. \ref{supp.fig:MinPhaseVar}.
    Both the phase variance of the measured state and of the coherent state exhibit the same trend and decrease for increasing photon numbers.
    This behavior of the phase variance has also been measured for coherent states in Ref. \cite{SBCR93}.
    Note that our measured phase variance is slightly higher than the one of the coherent state which can be attributed to the finite width of the selected phase-space region and to a lower amount of data in some of these regions.
